\documentclass[a4paper,10pt,superscriptaddress,pra,twocolumn,showpacs]{revtex4}
\usepackage{amssymb}
\usepackage{amsmath}
\usepackage{float}
\usepackage{graphicx}
\usepackage{pstricks}
\usepackage{pst-grad}
\usepackage{pst-node}
\usepackage{pst-coil}
\usepackage{multido}
\usepackage{pst-plot}
\usepackage{psfrag}
\usepackage{float}
\usepackage{mathrsfs}

\newcommand{\bra}[1]{\ensuremath{\langle#1|}}
\newcommand{\ket}[1]{\ensuremath{|#1\rangle}}

\newcommand{\tr}{\mathrm{tr}}

\newcommand{\mc}{\mathcal}
\newcommand{\therm}{\mathrm{th}}

\newcommand{\set}{\mathfrak C}
\newcommand{\proof}{\emph{Proof}}

\newcommand{\RM}{\Re M}
\newcommand{\IM}{\Im M}
\newcommand{\dmax}{d}
\newcommand{\pure}{\textrm{pure}}

\usepackage[latin1]{inputenc}
\usepackage{amsfonts}
\usepackage{amsmath}
\usepackage{amssymb}
\SpecialCoor

\definecolor{verd}{rgb}{0, .6, 0}

\begin{document}

\title{Information geometry of Gaussian channels}

\author{Alex Monras}

\affiliation{Dipartimento di Matematica e Informatica,
Universit\`{a} degli Studi di Salerno, Via Ponte don Melillo,
I-84084 Fisciano (SA), Italy}

\affiliation{CNR-INFM Coherentia, Napoli, Italy; CNISM Unit\`{a} di
Salerno; and INFN Sezione di Napoli, Gruppo collegato di Salerno,
Baronissi (SA), Italy}

\author{Fabrizio Illuminati}

\affiliation{Dipartimento di Matematica e Informatica,
Universit\`{a} degli Studi di Salerno, Via Ponte don Melillo,
I-84084 Fisciano (SA), Italy}

\affiliation{CNR-INFM Coherentia, Napoli, Italy; CNISM Unit\`{a} di
Salerno; and INFN Sezione di Napoli, Gruppo collegato di Salerno,
Baronissi (SA), Italy}

\affiliation{ISI Foundation for Scientific Interchange, Villa
Gualino, Viale Settimio Severo 65, I-10133 Torino, Italy}

\affiliation{Corresponding author. Electronic address:
illuminati@sa.infn.it}

\date{March 9, 2010}

\begin{abstract}
We define a local Riemannian metric tensor in the manifold of Gaussian channels and the distance that it induces. We adopt an information-geometric approach and define a metric derived from the Bures-Fisher metric for quantum states. The resulting metric inherits several desirable properties from the Bures-Fisher metric and is operationally motivated from distinguishability considerations: It serves as an upper bound to the attainable quantum Fisher information for the channel parameters using Gaussian states, under generic constraints on the physically available resources. Our approach naturally includes the use of entangled Gaussian probe states. We prove that the metric enjoys some desirable properties like stability and covariance. As a byproduct, we also obtain some general results in Gaussian channel estimation that are the continuous-variable analogs of previously known results in finite dimensions. We prove that optimal probe states are always pure and bounded in the number of ancillary modes, even in the presence of constraints on the reduced state input in the channel. This has experimental and computational implications: It limits the complexity of optimal experimental setups for channel estimation and reduces the computational requirements for the evaluation of the metric: Indeed, we construct a converging algorithm for its computation. We provide explicit formulae for computing the multiparametric quantum Fisher information for dissipative channels probed with arbitrary Gaussian states, and provide the optimal observables for the estimation of the channel parameters (e.g. bath couplings, squeezing, and temperature).

%
\end{abstract}


\maketitle

\section{Introduction}\label{sec:intro}
The theory of quantum channels provides a broad conceptual and mathematical framework to describe physical transformations on quantum states. Progress in quantum information technology is bringing long-standing questions related to quantum channels to the front line of research. Topics such as dissipation-assisted quantum computation~\cite{palma_quantum_1996}, quantum teleportation~\cite{bennett_teleportingunknown_1993}, quantum memories~\cite{fleischhauer_quantum_2002} and quantum state engineering~\cite{cirac_preparation_1994,dellannophysrep_2006} all have in common that they deal with quantum channels in one way or another.
%
One major question that has recently received significant amount of attention is the definition of a distance among quantum channels~\cite{aharonov_quantum_1998,gilchrist_distance_2005,caves_fidelity_2004}. The main motivation for such notion is the identification of a \emph{gold standard}~\cite{caves_fidelity_2004} against which all quantum processes could be compared, which would unify and systematize the way in which errors are treated and quantified. In~\cite{gilchrist_distance_2005} a reasonable, physically motivated set of requirements for such distance was introduced and some particular cases fulfilling most of them were identified. However, despite recent progress in the field~\cite{watrous_semidefinite_2009}, many questions remain open.

A similarly motivated program, as developed between the late sixties and the early nineties, regarded the problem of defining distances among quantum states.
The resulting theory is today well encompassed within the framework of Information Geometry~\cite{amari_methods_2001, petz_quantum_2007,bengtsson_geometry_2006}. In particular, this approach addresses questions regarding the distinguishability and the estimation of sets of quantum states. These questions led to the notions of quantum fidelity~\cite{Jozsa94}, Bures distance and quantum Fisher information (QFI) \cite{bures_extension_1969, helstrom_quantum_1976, holevo_probabilistic_1982, uhlmann_density_1993, hubner_explicit_1992, braunstein_statistical_1994}. As a result, the manifold of quantum states is endowed with a local Riemannian metric tensor which is physically motivated and serves as the gold standard for comparing quantum states. Additionally, the metric allows one to define the Bures distribution~\cite{slater_quantum_1996,bengtsson_geometrical_2005,bengtsson_geometry_2006}, the quantum analog to the Jeffreys' prior~\cite{jeffreys_invariant_1946}, which, among other applications, provides an operationally motivated prior distribution for Bayesian tomography and estimation techniques~\cite{bernardo_bayesian_2000}.

%



There exists in the literature a number of proposals for defining a distance for quantum channels. Most notably, the \emph{Jamiolkowski process distance}~\cite{gilchrist_distance_2005}, based on the Choi-Jamiolkowski isomorphism~\cite{jamiolkowski_72}, and the \emph{completely bounded trace norm} \cite{kitaev_97,gilchrist_distance_2005} establish a distance between two channels $\mc S_1$ and $\mc S_2$ by considering the distance $d(\rho_1,\rho_2)$ between the respective images of an arbitrary probe state $\rho$ under the action of the channels $\rho_i=(\mc S_i\otimes\mc I)\rho$. Such distance is then maximized over all possible probe states $\rho$ in order to obtain a fundamental measure of distinguishability between $\mc S_1$ and $\mc S_2$. The completely bounded trace norm is arguably the most appropriate choice both for physically motivated reasons and practical considerations: Indeed, it has been recently shown that it can be computed in polynomial time in the dimension of the system upon which the channels act~\cite{watrous_semidefinite_2009}. Despite this and other significant contributions, existing distance measures are inadequate to address a number of relevant situations. In particular, defining a distance on a manifold instead of a metric tensor hinders the task of defining natural prior distributions over the set of channels. Moreover, most of the existing distances are hard, if not impossible, to compute in infinite-dimensional systems such as continuous variables~\cite{wang_quantum_2007,adessoilluminati_2007}. Additionally, in a number of situations, arbitrarily good distinguishability between any two infinite-dimensional channels can be achieved if one allows for a large enough amount of resources~\cite{footnote1}, thus rendering the direct approach of optimization~\cite{watrous_semidefinite_2009} useless, unless some restrictions on the resources (regularization) are enforced.

In the present work we address the problem of defining a metric tensor on the set of $n$-mode 
Gaussian channels $\set$~\cite{lindblad_cloningquantum_2000}, and we study the ensuing physical consequences. The set of Gaussian channels can be regarded as a manifold once a proper parametrization $X:\mc O\rightarrow \set$ is established, where $\mc O\subseteq \mathbb R^d$ is an open subset in a $d$-dimensional real vector space. The set $\set=\{\mc S(X)\}$  can be equipped with a Riemannian metric following the spirit of the statistical distance~\cite{wootters_statistical_1981} and the Bures distance. Considering any two infinitesimally close quantum channels $\mc S_1\equiv\mc S(X)$ and $\mc S_2\equiv\mc S(X+dX)$, a metric tensor $\mathfrak J$ provides the infinitesimal distance between them as
\begin{equation}\label{eq:dsq}
	d^2(\mc S_1,\mc S_2)=\mathfrak J_{\mu\nu}(X)dX^\mu dX^\nu
\end{equation}
where $\mathfrak J(X)\geq0$. We use Einstein's summation convention, \emph{i.e.,} an index appearing once as a sub- and once as a super-index is automatically contracted.

As pointed out in~\cite{gilchrist_distance_2005} some physically motivated requirements should be imposed. Namely, the metric should meet the following criteria:
\begin{itemize}
	\item \emph{Stability}: The metric should be invariant under the addition of ancillary modes, namely $d^2(\mc S_1,\mc S_2)=d^2(\mc S_1\otimes\mc I,\mc S_2\otimes\mc I)$, where $\mc I$ corresponds to the identity channel on an arbitrary number of ancillary modes.
	\item \emph{Measurability}: This amounts to say that, once the metric is defined, the channel parameters can be determined by experimental means.
	\item \emph{Computability}: The metric $\mathfrak J(X)$ should be computable. This requirement is obviously unclear for the existing proposals when addressing infinite-dimensional channels.
	\item \emph{Physical meaning}:  The metric should have a clear operational meaning. This is certainly the case if the metric is derived from distinguishability considerations.
\end{itemize}
Other requirements (symmetry, non-degeneracy, and the triangle inequality) are guaranteed by any distance stemming from a metric tensor. The chaining condition imposed on~\cite{gilchrist_distance_2005} is not immediate to translate into the metric approach, and we will not address it in the present work. 
Finally, a purely formal requirement is in order for any well-defined metric tensor
\begin{itemize}
	\item \emph{Covariance}: The metric tensor has to transform covariantly under a reparametrization. Namely, if we perform the change $X\rightarrow X'(X)$, with $dX'^\mu=\Delta^\mu_{~\nu}(X) d X^\nu$ the metric tensor has to transform as $\mathfrak J_{\mu\nu}(X)=\mathfrak J'_{\lambda\sigma}(X')\Delta^\lambda_{~\mu}(X) \Delta^\sigma_{~\nu}(X)$ so that the distance $d(\mc S_1,\mc S_2)$ is invariant under reparametrization.
\end{itemize}

As mentioned earlier, some physical constraint on the probe states needs to be imposed in order to guarantee that the metric is well defined. We assume that this constraint is given in the form of a real-valued function $\phi(\rho)\leq\phi^\star$, where the specific form of $\phi$ is not of particular relevance. This function may have a practical motivation, and thus be chosen according to technical considerations, or may be used to regularize the divergencies that appear when resources are unlimited; we will refer to it as a \emph{resource budget}. From now on we tacitly assume that all quantities we define depend implicitly on the given choice of $\phi$.

The paper is organized as follows. Section \ref{sec:QFI} introduces the Bures-Fisher Metric, a concept central to our work, together with some of its properties. Section \ref{sec:metric} defines our channel metric and derives some of its main properties, namely covariance and stability, as well as showing that it can be computed with arbitrary precision, provided that one can compute Quantum Fisher information (QFI) matrices for all Gaussian probe states. In doing so, we generalize some results known previously for finite-dimensional channels into the constrained Gaussian setup. In Section \ref{sec:formulae} we derive explicit formulae for obtaining the QFI matrices for dissipative channels. Section \ref{sec:conclusions} provides some remarks and stresses the main questions left open, as well as the near-future applications of our results to the estimation of relevant channel properties such as bath couplings, temperature, and squeezing. The mathematical details are reported in five technical appendices.

\section{The Bures-Fisher metric}
\label{sec:QFI}

The main motivation for defining a metric in the manifold of Gaussian channels $\set$ is to obtain a parametrization-independent measure of distinguishability on $\set$. By expressing distances by means of a metric tensor we also obtain a notion of \emph{density} of channels. Such density thus provides a measure of \emph{how many} different channels can be distinguished in a neighborhood of a point $\mc S(X)$ with a given amount of resources $\phi^\star$. Such distinguishability-derived density is captured by the Jeffrey's prior~\cite{jeffreys_invariant_1946} in the case of probability distributions and the Bures prior~\cite{bures_extension_1969, hubner_explicit_1992} for the set of quantum states. These densities play a central role in the theory of Bayesian estimation~\cite{bernardo_bayesian_2000}.

Given two channels $\mc S_1$ and $\mc S_2$ and a specific probe state $\rho_0$, the maximal statistical distance~\cite{wootters_statistical_1981} attainable by any quantum measurement between the states resulting from the action of the channels, $\rho=\mc S_1 \rho_0$ and $\sigma=\mc S_2\rho_0$ is given by the Bures distance~\cite{bures_extension_1969, Jozsa94}
\begin{equation}
	d^2_B(\rho,\sigma)=2\left(1-\sqrt{F(\rho,\sigma)}\right),
\end{equation}
where
\begin{equation}
	F(\rho,\sigma)=\tr\sqrt{\sqrt\rho\,\sigma\sqrt\rho}
\end{equation}
is the quantum Fidelity. For infinitesimally close channels, $\mc S_1=\mc S(X)$ and $\mc S_2=\mc S(X+dX)$ the Bures distance can be expressed as,
\begin{equation}\label{eq:dbures}
	d^2_B(\rho,\sigma)=\frac{1}{4}\mathcal J(X|\rho_0)_{\mu\nu}dX^\mu dX^\nu
\end{equation}
where $\mc J(X|\rho_0)$ is the quantum Fisher information matrix (QFI) of the model $\{\rho(X)=\mc S(X)\rho_0\}$~\cite{helstrom_quantum_1976, holevo_probabilistic_1982, uhlmann_density_1993, braunstein_statistical_1994}. The QFI matrix can be computed from the \emph{symmetric logarithmic derivatives} (SLD) $\Lambda_\mu$, which are the Hermitian operators that satisfy the equation
\begin{equation}
	\label{eq:SLDdef}
	\partial_\mu \rho(X)=\Lambda_\mu(X)\circ \rho(X),
\end{equation}
where we have introduced the symmetric product for operators, $A\circ B=\frac{1}{2}(AB+BA)$. The QFI is then
\begin{equation}
	\label{eq:QFIdef}
	\mc J_{\mu\nu}(X)=\tr[\rho \,\Lambda_\mu\circ\Lambda_\nu],
\end{equation}
where we have dropped the explicit $X$ dependency in $\rho$ and $\Lambda_\mu$. The SLD's play an important role in the theory of quantum inference~\cite{braunstein_statistical_1994}.

The QFI has been reviewed a number of times in the literature, from which we emphasize~\cite{bengtsson_geometry_2006,Gill01asymptoticsin, petz_quantum_2007,Paris_Estimation_2009}. It enjoys several useful properties
\begin{itemize}
	\item The SLD has zero expectation,
	\begin{equation}\label{eq:consistency1}
		\tr[\rho \Lambda_\mu]=\tr[\rho \circ\Lambda_\mu]=\partial_\mu\tr\rho=0.
	\end{equation}
	\item The QFI is real, symmetric and positive semi-definite,
	\begin{equation}
		\theta^\mu \mc J_{\mu\nu}(X) \theta^\nu=\tr[\rho\,(\theta^\mu \Lambda_\mu)^2]\geq0
	\end{equation}
	which follows from $\rho\geq0$ and $(\theta^\mu\Lambda_\mu)^2\geq0$.
	\item The SLD and the QFI are covariant quantities. Given a parameter $\theta(X)$ such that $\partial_\theta\rho(X)=\theta^\mu \partial_\mu\rho(X)$ we can define the SLD associated to $\theta$, $\Lambda_\theta=\theta^\mu\Lambda_\mu$. We can equally define the QFI associated to $\theta$ as $\mathcal J_\theta=\theta^\mu \mc J_{\mu\nu}\theta^\nu$.
%
%
	\item The QFI is monotonic under completely-positive trace-preserving (CPTP) maps. Writing $\mc J(\rho)_{\mu\nu}=\tr[\rho\,\Lambda_\mu^\rho \circ \Lambda_\nu^\rho]$ we have
	\begin{equation}
		\mc J(\rho)\geq \mc J(\mc E\rho),
	\end{equation}
	where $\mc E$ is any CPTP map. In particular, if $\mc E$ is a unitary map, it holds that $\mc J(\rho)=\mc J(\mc E\rho)$.
\end{itemize}

Apart from these properties, the QFI plays a central role in the theory of quantum statistical inference by placing a lower bound on the attainable variance of any unbiased estimator of the parameters $X$~\cite{helstrom_quantum_1976,holevo_probabilistic_1982,braunstein_statistical_1994, Gill01asymptoticsin, Paris_Estimation_2009}.

\section{A channel metric}\label{sec:metric}

\psset{xunit=1,yunit=1}
\psset{linearc=.01,fillstyle=solid}

\definecolor{LightOrange}{cmyk}{0,0.2,0.4,0}%
\definecolor{DarkOrange}{cmyk}{0,0.2,0.8,0}%

\definecolor{MeasureCol1}{rgb}{.4,0.5,1}
\definecolor{MeasureCol2}{rgb}{.4,1,1}
\definecolor{SymplecticCol1}{rgb}{.9 1 .9}
\definecolor{SymplecticCol2}{rgb}{.4 1 .6}
\definecolor{ThermalCol1}{rgb}{1 0 0}
\definecolor{ThermalCol2}{rgb}{1 1 .4}

\def\channel{%
	\pspolygon[linearc=.2,fillstyle=gradient, gradmidpoint=1,gradangle=45,gradbegin=LightOrange,gradend=DarkOrange](0,-.5)(2,-.5)(2,.5)(0,.5)%
	\rput(1,0){$\mc S(X)$}
	}

\def\symplectic#1(#2){%
\pspolygon[fillcolor=cyan,linearc=.2, fillstyle=gradient,gradmidpoint=1,gradangle=45,gradbegin=SymplecticCol1,gradend=SymplecticCol2](0,-.5)(1,-.5)(! 1 -.5 #2 add)(! 0 -.5 #2 add)%
	\rput(!.5 -.5 .5 #2 mul add){#1}%
	}

\def\thermalTm(#1,#2){%
\psset{fillstyle=gradient,gradmidpoint=1,gradangle=0,gradbegin=ThermalCol1,gradend=ThermalCol2}
	\rput(!#1 .5 add #2 .15 sub){\rotateleft{\pscustom{\parabola(-.1,0)(0,.2)%
	\closepath
	}}}
	\rput(!#1 .5 add #2){\rotateleft{\pscustom{\parabola(-.1,0)(0,.2)%
	\closepath
	}}}
	\rput(!#1 .5 add #2 .15 add){\rotateleft{\pscustom{\parabola(-.1,0)(0,.2)%
	\closepath
	}}}
}

\def\fiberTm(#1,#2,#3){%
	\psline(!#1 #3 -.15 add)(!#2 #3 -.15 add)%
	\psline(#1,#3)(#2,#3)%
	\psline(!#1 #3 .15 add)(!#2 #3 .15 add)%
}

\def\thermalFm(#1,#2){%
	\psset{fillstyle=gradient,gradmidpoint=1,gradangle=45,gradbegin=ThermalCol1,gradend=ThermalCol2}
	\rput(!#1 .5 add #2 .2 sub){\rotateleft{\pscustom{\parabola(-.1,0)(0,.2)%
	\closepath
	}}}
	\rput(!#1 .5 add #2 .1 sub){\rotateleft{\pscustom{\parabola(-.1,0)(0,.2)%
	\closepath
	}}}
	\rput(!#1 .5 add #2){\rotateleft{\pscustom{\parabola(-.1,0)(0,.2)%
	\closepath
	}}}
	\rput(!#1 .5 add #2 .1 add){\rotateleft{\pscustom{\parabola(-.1,0)(0,.2)%
	\closepath
	}}}
	\rput(!#1 .5 add #2 .2 add){\rotateleft{\pscustom{\parabola(-.1,0)(0,.2)%
	\closepath
	}}}
}

\def\fiberFm(#1,#2,#3){%
	\psline(!#1 #3 -.2 add)(!#2 #3 -.2 add)%
	\psline(!#1 #3 -.1 add)(!#2 #3 -.1 add)%
	\psline(#1,#3)(#2,#3)%
	\psline(!#1 #3 .1 add)(!#2 #3 .1 add)%
	\psline(!#1 #3 .2 add)(!#2 #3 .2 add)%
}

\def\gap{.08 }
\def\measure(#1,#2,#3){%
\psellipticarc[showpoints=true,fillstyle=gradient,gradmidpoint=1,gradangle=45,gradbegin=MeasureCol1,gradend=MeasureCol2](! #1 -.5 #2 #3 2 div add add)(!1 #3 2 div \gap sub){-90}{90}
	\psline(!#1 #2 -.5 add \gap add)(! #1 -.5 #2 #3 add add \gap sub)
	}

In this section we define our proposed channel metric and study its properties.
%
%
Analogously to Eq.~\eqref{eq:dbures}, we expect that the distance between neighboring channels can be expressed as in Eq.~\eqref{eq:dsq}. This is a requirement that is not met by all distance measures, and it is yet unclear that a straightforward extrapolation of the statistical distance applied to channels will fulfill such expectation. Thus, instead of focusing on maximizing the statistical distance, we will search for a metric tensor that provides an upper bound to it. Namely,
\begin{enumerate}
	\item [1)] $\mathfrak J(X)$ should provide an upper bound to the Bures distance between states attainable within some resource budget $\phi^\star$.
\end{enumerate}
Following these considerations, and sticking to our definition of $\rho$ and $\sigma$, we impose the following condition on $\mathfrak J(X)$
\begin{equation}
	 d_B^2(\rho,\sigma)\leq \frac{1}{4}\mathfrak J(X)_{\mu\nu}dX^\mu dX^\nu
\end{equation}
for any possible $dX$ and any chosen probe state $\rho_0$ fulfilling $\phi(\rho_0)\leq\phi^\star$. Imposing this for any $dX$ implies that
\begin{equation}\label{eq:upperbound}
	\mc J(X|\rho_0)\leq \mathfrak J(X)\quad \forall \rho_0~s.t.~\phi(\rho_0)\leq\phi^\star.
\end{equation}
%
Notice, however, that this condition 
is not sufficient to uniquely specify $\mathfrak J(X)$. In general, one would expect that the distance established between two points is not unnecessarily large. However, a tight bound for $d_B(X,X+dX)$ for any $dX$ may not be expressible the form of Eq.~\eqref{eq:dsq}. This is because minimizing the distance between a pair of points $\mc S(X)$ and $\mc S(X+dX)$ may not automatically minimize the distance between $\mc S(X)$ and another $\mc S(X+dX')$. This can be formalized as follows. Let $C$ be the set of achievable QFI matrices for a given channel and resource budget $\phi^\star$. There may exist several matrices which are tight upper bounds to $C$, namely, that they are upper bounds, and no smaller matrix exists which is also an upper bound. This is a direct consequence of the partial-ordered nature of matrices. There is, however, a natural way to precisely specify $\mathfrak J(X)$ while respecting all the above criteria. One expects that
\begin{enumerate}
	\item [2)] The volume element specified by $\mathfrak J(X)$ should be minimal.
\end{enumerate}
 This translates into imposing that $\sqrt{\det \mathfrak J(X)}$ is minimal.

Thus, our metric corresponds to an upper bound on the attainable QFI for any set of Gaussian quantum channels $\set$, tested with Gaussian probe states fulfilling some regularization condition $\phi(\rho_0)\leq\phi^\star$, and providing the smallest possible volume element in the manifold of channels. Restricting to the Gaussian domain yields a manageable parametrization of the probe states and allows to obtain explicit formulae for the QFI. In addition, the geometry induced by our metric will have an immediate practical interpretation. It is a bound to the distinguishability attainable by Gaussian states. Moreover, recent analysis~\cite{monras_illuminati_prep} suggests that entangled Gaussian states are as sensitive as single-mode de-Gaussified states~\cite{illuminati_salerno_heisenberg-limited_Fock_2008} in quantum statistical inference. This claim is yet to be proven.

As a first approach to the problem of defining a channel metric it is relevant to discuss the Jamiolkowski distance introduced in~\cite{gilchrist_distance_2005}. It is well known that the Jamiolkowski isomorphism~\cite{jamiolkowski_72} can be used to encode a $d$-dimensional quantum channel $\mc S$ into a $d^2$-dimensional quantum state $\sigma_{\mc S}=(\mc S\otimes\openone)\Phi$, where $\Phi$ is a maximally entangled state. One may consider that distinguishability between $\sigma_{\mc S_1}$ and $\sigma_{\mc S_2}$ be related to the distinguishability between $\mc S_1$ and $\mc S_2$. The state $\Phi$, however, is not likely to comply with any physically motivated resource restriction $\phi$. Instead, one must consider the attainable distinguishability within the limits imposed by $\phi$.

We proceed, in the next subsections, to formally define the metric tensor $\mathfrak J(X)$ based on the previously announced criteria 1) \& 2), and prove the covariance, stability and computability properties.

\subsection{Defining $\mathfrak J(X)$}

\begin{figure}[b]
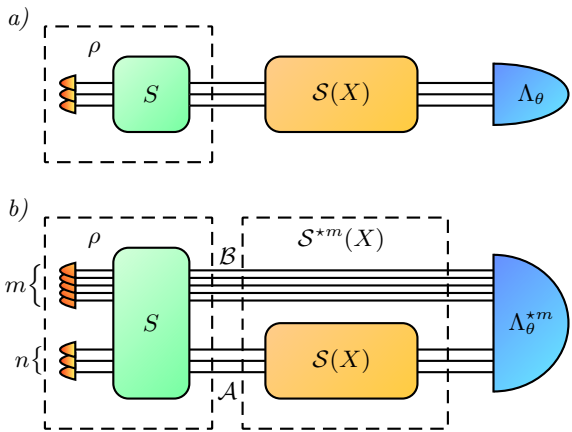

\begin{center}
\pspicture(-.5,0)(7.5,2.5)
\rput(-.25,2){\emph{a)}}
\rput(1,1){\symplectic$S$(1)}
\fiberTm(2,6,1)
\rput(3,1)\channel
\measure(6,1,1)
\thermalTm(0,1)
\fiberTm(.5,1,1)
\pspolygon[fillstyle=none,linestyle=dashed](.1,.1)(2.3,0.1)(2.3,1.9)(.1,1.9)
\rput(.75,1.6){$\rho$}
\rput(6.5,1){$\Lambda_\theta$}
\endpspicture\\%
\pspicture(-.5,0)(7.5,3.5)
\rput(-.25,3){\emph{b)}}
\fiberTm(.5,6,1)
\fiberFm(.5,6,2)
\rput(1,1){\symplectic$S$(2)}
\rput(3,1)\channel
\measure(6,1,2)
\thermalTm(0,1)
\thermalFm(0,2)
\rput(-0.1,1){$n\big\{$}
\rput(-0.175,2){$m\Big\{$}
\pspolygon[fillstyle=none,linestyle=dashed](.1,.1)(2.3,0.1)(2.3,2.9)(.1,2.9)
\rput(.75,2.6){$\rho$}
\pspolygon[fillstyle=none,linestyle=dashed](2.7,0.1)(5.4,0.1)(5.4,2.9)(2.7,2.9)
\rput(4,2.6){$\mc S^{\star m}(X)$}
\rput(6.5,1.5){$\Lambda_\theta^{\star m}$}
\rput(2.5,.6){$\mc A$}
\rput(2.5,2.4){$\mc B$}

\endpspicture\\%
\end{center}
\caption{\label{fig:onemode}
(Color online)~Measurement schemes \emph{a)} without ancillas and \emph{b)} with ancillas, combined with the most general possible measurement. Scheme \emph{b)} includes \emph{a)} and hence, must be equally or more efficient than \emph{a)}.}
\end{figure}

Given a parameterized family of $n$-mode channels $\set=\{\mc S(X)\}$, let us define the {\em quantum model} $\mc M[\set|\rho]=\{\rho(X)=\mc S(X){\rho_0}\}$, {\em i.e.} the parameterized set of all possible quantum states resulting from an initial probe ${\rho_0}$, under the action of the set of channels $\set$. The associated QFI is
\begin{equation}
	\mc J_{\mu\nu}(\mc S(X)|{\rho_0})=\tr[\rho(X)~\Lambda_\mu\circ\Lambda_\nu],
\end{equation}
where $\Lambda_\mu$ are the associated SLD's. This corresponds to the situation depicted in Fig.~\ref{fig:onemode}\emph{a}.

Consider next the \emph{$m$-completed channel} as $\mc S^{\star m}(X)=\mc S(X)\otimes\mc I_m$ where $\mc I_m$ represents the identity channel on $m$ ancillary modes. We thus obtain the \emph{$m$-completed set} $\set^{\star m}=\{\mc S^{\star m}(X)\}$. Given an $(n+m)$-mode probe state ${\rho}$ we can define a new quantum model, $\mc M[\set^{\star m}|{\rho_0}]=\{\rho^{\star m}(X)=\mc S^{\star m}(X){\rho_0}\}$. The QFI for the new model is
\begin{equation}
	\mc J_{{\mu\nu}}(\mc S^{\star m}(X)|{\rho_0})=\tr[\rho^{\star m}(X)~\Lambda^{\star m}_\mu\circ\Lambda^{\star m}_\nu],
\end{equation}
where $\rho^{\star m}(X)=\mc S^{\star m}(X){\rho_0}$ is the probe state under the action of the $m$-completed channel, and the SLD's $\Lambda^{\star m}_\mu$ obviously correspond to the model $\mc M[\set^{\star m}|{ \rho_0}]$. The extension of $\set$ to $\set^{\star m}$ is depicted in Fig.~\ref{fig:onemode}\emph{b}.

Now we turn to the constraint $\phi$. Let us denote the $n$ modes upon which the channels $\mc S(X)$ act as $a_k, k\in\{1,\ldots,n\}$, and collectively referred to as $\mc A$ while the ancillary modes are denoted by $b_k, k\in\{1,\ldots,m\}$, collectively denoted $\mc B$. We may find it useful to consider yet another set of modes, collectively denoted as $\mc C$. We define the \emph{acceptable} constraints, as those that only involve the reduced density operator in modes $\mc A$, $\rho_{\mc A}=\tr_{\bar{\mc A}}\rho$, namely
\begin{equation}
	\phi(\rho)\equiv\phi(\rho_{\mc A}).
\end{equation}
This {choice} is physically motivated by the fact that relevant resources should only involve degrees of freedom accessible to $\mc A$. A natural choice for the function $\phi$ could be the average photon number in $\mc A$, $\phi(\rho)=\tr[\rho \sum_k a_k^\dagger a_k]$, which has been used previously as a comparison reference for several estimation problems. Nevertheless, our results are general and not restricted to this particular choice.

{Finally let $C(\mc S(X)|\phi^\star)=\{ \mc J(\mc S(X)|\rho)~|~ \phi(\rho)\leq\phi^\star\}$ be the set of all QFI's achievable by any Gaussian state fulfilling the constraint $\phi(\rho)\leq\phi^\star$, and $C_\pure(\mc S(X)|\phi^\star)=\{ \mc J(\mc S(X)|\rho)~|~ \phi(\rho)\leq\phi^\star,\rho^2=\rho\}$ the corresponding set when restricting to pure states}. Let $C$ be any set of positive semidefinite bounded matrices, and $\mc J\in C$. Define the function $M(\mc J)=\{j\geq\mc J\}$ to be the set of all upper bounds to $\mc J$, and $M(C)=\cap_{\mc J\in C} M(\mc J)$ be the set of all upper bounds common to all matrices in $C$.

We are now in the position to introduce the metric for the set $\set$. Let $\mc S$ be a channel in $\set$. The metric at point $\mc S$ is
\begin{equation}\label{eq:definition}
	\mathfrak J(\mc S)=\underset{\{j\in M(C({\mc S^{\star n}}|\phi^\star))\}}{\arg \inf} \det j.
\end{equation}
That is, the matrix with smallest determinant that is greater or equal to all possible QFI's achievable by any $2n$-mode Gaussian probe state fulfilling the constraint $\phi(\rho)\leq\phi^\star$. This is a positive semidefinite matrix and certainly qualifies as a metric. It is measurable, in the sense that it is a function of the parameters $X$ which are themselves measurable [see Sec.~\ref{sec:measure}]. Notice that in very pathological cases, there may be more than one solution to the minimization problem. These are highly symmetric and unlikely situations which we will not deal with.  We dedicate the following subsections to discuss the most relevant properties of $\mathfrak J$, and provide an approximation method to compute it.

The idea behind this definition is to provide an upper bound to the achievable Fisher information for any parameter of interest, under the constraint $\phi(\rho)\leq\phi^\star$. While any matrix greater than those in $C({\mc S^{\star n}}|\phi^\star)$ would certainly qualify as an upper bound, the partial-ordered nature of matrices prevents from having, in the general case, a well defined supremum. Instead, the minimization of the determinant in Eq.~\eqref{eq:definition} aims at reducing the \emph{volume element} to a minimum while still providing an upper bound to the attainable Fisher information.


\subsection{Covariance}\label{sec:covariance}
Notice that under a reparametrization $X\rightarrow X'(X)$ we have
\begin{equation}
	\label{eq:covariantSLD}
	\Lambda_\mu\rightarrow \Lambda_{\mu'}=\Lambda_\mu\Delta^\mu_{~\mu'},
\end{equation}
where $\Delta^\mu_{~\mu'}\equiv{\partial X^\mu/\partial {X'}^{\mu'}}$, where primed indices refer to the new coordinates and unprimed indices correspond to the old coordinates, as is customary in the notation of General Relativity. Eq.~\eqref{eq:covariantSLD} in turn implies that the QFI is a covariant quantity
\begin{equation}
	\mc J_{\mu\nu}\rightarrow \mc J'_{\mu'\nu'}=\mc J_{\mu\nu} \Delta^{\mu}_{~\mu'} \Delta^{\nu}_{~\nu'},
\end{equation}
Let us denote this transformation law in shorthand notation $\mc J\rightarrow \Delta^T \mc J \Delta$.
All elements in $C({\mc S^{\star n}}(X)|\phi^\star)$ transform in the same way under a given reparametrization. Accordingly, the set $M(C({\mc S^{\star n}}(X)|\phi^\star))$
also transforms covariantly. Namely, for any matrix $j\in M(C({\mc S^{\star n}}(X)|\phi^\star))$ we have
\begin{equation}
	\Delta^T j \Delta\in M(C({\mc S^{\star n}}(X')|\phi^\star)),
\end{equation}
and viceversa.

Finally the determinants of all elements in $M(C(\mc S(X')|\phi^\star))$ are just those of $M(C(\mc S(X)|\phi^\star))$ multiplied by a factor $\det \Delta^T \Delta$. Therefore, assuming that under parametrization $X$ the minimum was achieved for $\mathfrak J$, the new minimum under parametrization $X'$ will be achieved by $\mathfrak J'=\Delta^T \mathfrak J \Delta$ [$\mathfrak J'_{\mu'\nu'}=\mathfrak J_{\mu\nu} \Delta^{\mu}_{~\mu'}\Delta^{\nu}_{~\nu'}$], {as required}.

\subsection{Stability}\label{sec:stability}
The stability requirement has been discussed previously in the
literature~\cite{aharonov_quantum_1998,gilchrist_distance_2005}. We
dedicate this section to prove that $\mathfrak J(\mc S{})$ is invariant
under the addition of ancillas, namely
\begin{equation}
    \mathfrak J(\mc S{}\otimes \mc I)=\mathfrak J(\mc S{}),
\end{equation}
where $\mc I$ is the identity channel on an arbitrary number $n'$ of
modes. We remark that while adding ancillary modes, we still require the
constraints $\phi$ to act on the original set of modes. Observe that
\begin{equation}\label{eq:extendedmetric}
    \mathfrak J(\mc S{}\otimes \mc I)=\underset{\{j\in M(C(\mc S^{\star
(n+2n')}|\phi^\star))\}}{\arg \inf} \det j.
\end{equation}

Observe that if $X\subseteq Y$ then $M(Y)\subseteq M(X)$. Given two sets of
positive semidefinite matrices $X$ and $Y$, we say $X\succeq  Y$ if and
only if for
any $\forall y\in Y~\exists x\in X$ such that $x\geq y$. This binary
relation is reflexive ($X \succeq X$) and transitive ($X\succeq Y$ and
$Y\succeq Z\,\Rightarrow\,X\succeq Z$). {Also}
$X\succeq  Y\,\Rightarrow\,M(X)\subseteq M(Y)$. We will say that two sets are
\emph{equivalent} ($X\simeq Y$) if and only if $M(X) = M(Y)$.

Let $n$ be the number of modes upon which $\mc S$ acts. The proof
consists in two main steps.
\begin{enumerate}
    \item We will first show that the set of QFI matrices attainable
using \emph{pure} probe states with $m>n$ ancillary modes is equivalent
to those attainable with \emph{pure} probes using only $n$ ancillary modes.
    {\item We will then show that the set of QFI matrices attainable
with $(n+m)$-mode states ($m>n$) is equivalent to those attainable with \emph{pure}
$(n+m)$-mode states.}
\end{enumerate}

\begin{figure}[b]
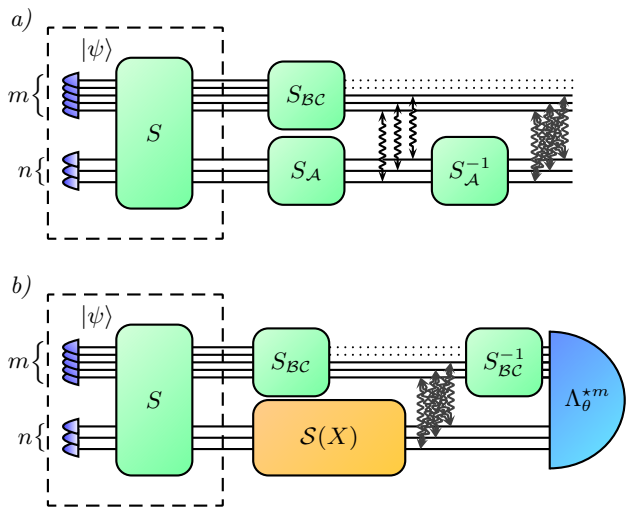

\psset{coilarm=.15,coilwidth=.1,linearc=.01}
\begin{center}
\pspicture(-.5,0)(7.5,3.5)
\rput(-.25,3){\emph{a)}}
\fiberTm(.5,7,1)
\fiberFm(.5,3,2)
\psline(4,2)(7,2)
\psline(4,1.9)(7,1.9)
\psline(4,1.8)(7,1.8)
\psline[linestyle=dotted,dotsep=2pt](4,2.1)(7,2.1)
\psline[linestyle=dotted,dotsep=2pt](4,2.2)(7,2.2)
\rput(1,1){\symplectic$S$(2)}
\rput(3,1.05){\symplectic$S_{\mc A}$(.9)}
\rput(5.15,1.05){\symplectic$S_{\mc A}^{-1}$(.9)}
\rput(3,2.05){\symplectic$S_{\mc{BC}}$(.9)}

\pszigzag{<->}(4.5,1.8)(4.5,0.85)
\pszigzag{<->}(4.7,1.9)(4.7,1)
\pszigzag{<->}(4.9,2)(4.9,1.15)

\psset{linecolor=darkgray}
\pszigzag{<->}(6.5,1.8)(6.5,0.85)
\pszigzag{<->}(6.7,1.9)(6.7,1)
\pszigzag{<->}(6.9,2)(6.9,1.15)
\pszigzag{<->}(6.5,1.8)(6.9,1.15)
\pszigzag{<->}(6.7,1.9)(6.5,0.85)
\pszigzag{<->}(6.9,2)(6.7,1)
\pszigzag{<->}(6.5,1.8)(6.7,1)
\pszigzag{<->}(6.7,1.9)(6.9,1.15)
\pszigzag{<->}(6.9,2)(6.5,0.85)
\psset{linecolor=black}

\thermalTm(0,1)
\thermalFm(0,2)
\pspolygon[fillstyle=none,linestyle=dashed](.1,.1)(2.4,0.1)(2.4,2.9)(.1,2.9)
\rput(.75,2.6){$\ket\psi$}
\rput(-0.1,1){$n\big\{$}
\rput(-0.175,2){$m\Big\{$}

\endpspicture\\%
\pspicture(-.5,0)(7.5,3.5)
\rput(-.25,3){\emph{b)}}
\fiberTm(.5,7,1)
\fiberFm(.5,3,2)
\psline(3.5,2)(7,2)
\psline(3.5,1.9)(7,1.9)
\psline(3.5,1.8)(7,1.8)
\psline[linestyle=dotted,dotsep=2pt](3.5,2.1)(7,2.1)
\psline[linestyle=dotted,dotsep=2pt](3.5,2.2)(7,2.2)
\rput(1,1){\symplectic$S$(2)}
\rput(2.8,2.05){\symplectic$S_{\mc{BC}}$(.9)}

\rput(5.6,2.05){\symplectic$S_{\mc{BC}}^{-1}$(.9)}

\fiberFm(6.6,6.7,2)
\psset{linecolor=darkgray}
\pszigzag{<->}(5,1.8)(5,0.85)
\pszigzag{<->}(5.2,1.9)(5.2,1)
\pszigzag{<->}(5.4,2)(5.4,1.15)
\pszigzag{<->}(5,1.8)(5.4,1.15)
\pszigzag{<->}(5.2,1.9)(5,0.85)
\pszigzag{<->}(5.4,2)(5.2,1)
\pszigzag{<->}(5,1.8)(5.2,1)
\pszigzag{<->}(5.2,1.9)(5.4,1.15)
\pszigzag{<->}(5.4,2)(5,0.85)
\psset{linecolor=black}
\thermalTm(0,1)
\thermalFm(0,2)
\rput(-0.1,1){$n\big\{$}
\rput(-0.175,2){$m\Big\{$}

\pspolygon[fillstyle=none,linestyle=dashed](.1,.1)(2.4,0.1)(2.4,2.9)(.1,2.9)
\rput(.75,2.6){$\ket\psi$}

\rput(2.8,1)\channel
\measure(6.7,1,2)
\rput(7.2,1.5){$\Lambda_\theta^{\star m}$}

\endpspicture\\%

\end{center}
\caption{\label{fig:reduction}
(Color online)~\emph{a)} the Schmidt decomposition theorem ensures that
in an $(n+m)$-mode state ($m>n$), one can locally bring $m-n$ modes to the
vacuum. Therefore, for any pair $(\ket\psi,\Lambda_\theta)$ of $n+m$
modes there exists a pair $(S_{\mc{BC}} \ket\psi, S_{\mc{BC}}^\dagger
\Lambda_\theta S_{\mc{BC}})$ of $2n$-modes which performs equally well.
The remaining modes can be traced out of $S_{\mc{BC}}\ket\psi$ leading
to a generalized measurement, which can be described by a POVM. The
optimal POVM on the $2n$-mode state is given by the corresponding SLD,
$\Lambda_\theta^{\star n}$.}
\end{figure}

\noindent\emph{Proof of 1}: 
Consider an $(n+m)$-mode setup such as the one
in {Fig.~\ref{fig:onemode}\emph{b}} ($m>n$). The Schmidt decomposition-like
theorem~\cite{holevo_evaluating_2001,giedke_entanglement_2003,botero_modewise_2003,Serafini_localizable_2005}
states that, given an $(n+m)$-mode pure Gaussian state $\ket\psi_{\mc
A\mc B\mc C}$ separated into parties $\mc A$ ($n$ modes through the
channel), $\mc B$ ($n$ modes) and $\mc C$ ($m-n$ modes), one can always
reduce the system to $n$ pairwise squeezed states between $\mc A$ and
$\mc B$, and $m-n$ vacuum states in $\mc C$ by means of local symplectic
transformations on $\mc A$ ($S_{\mc A}$) and joint symplectic
transformations on $\mc B$ and $\mc C$ ($S_{\mc B \mc
C}$)~[Fig.~\ref{fig:reduction}\emph{a}]. Therefore, for any pure Gaussian
state $\ket\psi_{\mc A\mc B\mc C}$ there is another $2n$-mode state
$\ket{\psi_{2n}}_{\mc A\mc B}$ such that
\begin{equation}\label{eq:reducestate}
    S_{\mc B \mc C}\ket\psi_{\mc A\mc B\mc C}=\ket{\psi_{2n}}_{\mc A\mc
B}\ket0^{\otimes(n-m)}_{\mc C}.
\end{equation}
Defining $\mc E$ as $\mc E\rho= S_{\mc B\mc C} \rho S^\dagger_{\mc B A\mc C}$ and noting that $\mc S^{\star(n+m)}\mc E \rho=\mc E\mc S^{\star (n+m)}\rho$ we have, following from the unitary invariance of the QFI,
\begin{equation}
	\mc J(\mc S^{\star m}|\ket{\psi}_{\mc A\mc B\mc C})=\mc J(\mc S^{\star m}|\,\mc E\ket{\psi}_{\mc A\mc B\mc C}).
\end{equation}
Moreover, using $\mc S^{\star m}(\rho\otimes\ket{0}\bra0^{\otimes (m-n)})=(\mc S^{\star n}\rho)\otimes\ket{0}\bra0^{\otimes (m-n)}$ together with Eq.~\eqref{eq:reducestate} and the additivity of the QFI [obviously $\mc J(\mc I|\ket0^{\otimes (m-n)})=0$] we have
\begin{eqnarray}\nonumber
	\mc J(\mc S^{\star m}|\,\mc E\ket{\psi}_{\mc A\mc B\mc C})&=&\mc J(\mc S^{\star m}|\,\ket{\psi_{2n}}\!\otimes\!\ket0^{\otimes (n-m)})\\
	&=&\mc J(\mc S^{\star n}|\,\ket{\psi_{2n}}_{\mc A\mc B}),
\end{eqnarray}
while $\phi(\ket\psi)=\phi(\ket{\psi_{2n}})$. This shows that $C_{\pure}(\mc S^{\star m}|\phi^\star)\subseteq C_{\pure}(\mc S^{\star n}|\phi^\star)$, which combined with the trivial inverse inclusion yields
\begin{equation}
	C_{\pure}(\mc S^{\star m}|\phi^\star)= C_{\pure}(\mc S^{\star n}|\phi^\star),\quad\forall m>n
\end{equation}
and therefore
%
%
\begin{equation}\label{eq:pureequal}
    ~M(C_{\pure}(\mc S^{\star n}|\phi^\star))=
M(C_{\pure}(\mc S^{\star m}|\phi^\star)),\quad \forall m>n~~~\blacksquare
\end{equation}\nopagebreak[4]

This makes it clear that all $C_\pure(\mc S^{\star m}|\phi^\star)$ with $m\geq n$ are equivalent. This result has a similar flavor to the well known fact that maximally entangled states are optimal for estimation of $d$-dimensional unitary operations and several other finite-dimensional channels~\cite{optimal_entangled}, which means that a $d$-dimensional ancilla is sufficient.  In our case, maximal entanglement will in general be forbidden by the resource constraint $\phi^\star$, but nevertheless the ancilla need not be larger than the system itself. \\*

\definecolor{ThermalCol1}{rgb}{0 0 1}
\definecolor{ThermalCol2}{rgb}{1 1 1}

\definecolor{ThermalCol1}{rgb}{1 0 0}
\definecolor{ThermalCol2}{rgb}{1 1 .4}
\begin{figure}[b]
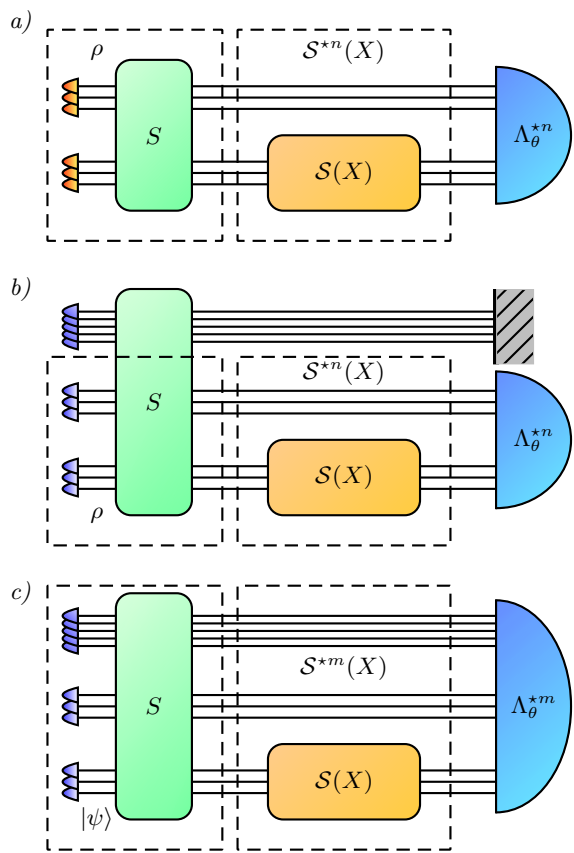

\begin{center}
\pspicture(-.5,0)(7.5,3.5)
\rput(-.25,3){\emph{a)}}
\fiberTm(.5,6,1)
\fiberTm(.5,6,2)
\rput(1,1){\symplectic$S$(2)}
\rput(3,1)\channel
\measure(6,1,2)
\rput(6.5,1.5){$\Lambda_\theta^{\star n}$}
\thermalTm(0,1)
\thermalTm(0,2)
\pspolygon[fillstyle=none,linestyle=dashed](.1,.1)(2.4,0.1)(2.4,2.9)(.1,2.9)
\rput(.75,2.6){$\rho$}

\pspolygon[fillstyle=none,linestyle=dashed](2.6,0.1)(5.4,0.1)(5.4,2.9)(2.6,2.9)
\rput(4,2.6){$\mc S^{\star n}(X)$}
\endpspicture\\%

\definecolor{ThermalCol1}{rgb}{0 0 1}
\definecolor{ThermalCol2}{rgb}{1 1 1}

\pspicture(-.5,0)(7.5,4)
\rput(-.25,3.5){\emph{b)}}
\fiberTm(.5,6,1)
\fiberTm(.5,6,2)
\fiberFm(.5,6,3)
\rput(1,1){\symplectic$S$(3)}
\rput(3,1)\channel
\measure(6,1,2)
\rput(6.5,1.5){$\Lambda_\theta^{\star n}$}
\thermalTm(0,1)
\thermalTm(0,2)
\thermalFm(0,3)
\pspolygon[fillstyle=none,linestyle=dashed](.1,.1)(2.4,0.1)(2.4,2.6)(.1,2.6)
\rput(.75,.5){$\rho$}

\pspolygon[fillstyle=none,linestyle=dashed](2.6,0.1)(5.4,0.1)(5.4,2.6)(2.6,2.6)
\rput(4,2.38){$\mc S^{\star n}(X)$}

\psline[linewidth=0.08](6,2.5)(6,3.5)
\psframe*[fillstyle=hlines,linecolor=lightgray](6,2.5)(6.5,3.5)
\psframe[linestyle=none,fillstyle=hlines](6,2.5)(6.5,3.5)

\endpspicture\\%
\pspicture(-.5,0)(7.5,4)
\rput(-.25,3.5){\emph{c)}}
\fiberTm(.5,6,1)
\fiberTm(.5,6,2)
\fiberFm(.5,6,3)
\rput(1,1){\symplectic$S$(3)}
\rput(3,1)\channel
\measure(6,1,3)
\rput(6.5,2){$\Lambda_\theta^{\star m}$}
\thermalTm(0,1)
\thermalTm(0,2)
\thermalFm(0,3)
\pspolygon[fillstyle=none,linestyle=dashed](.1,.1)(2.4,0.1)(2.4,3.6)(.1,3.6)
\rput(.75,.5){$\ket\psi$}

\pspolygon[fillstyle=none,linestyle=dashed](2.6,0.1)(5.4,0.1)(5.4,3.6)(2.6,3.6)
\rput(4,2.5){$\mc S^{\star m}(X)$}

\endpspicture\\%
\end{center}
\caption{\label{fig:purification}
(Color online)~Measurement schemes \emph{a)} with a thermal state in a
$2n$-mode setup, \emph{b)} Equivalent setup with a pure state containing
ancillary modes that are not measured, \emph{c)} The optimal measurement
including the ancillary modes. Obviously, the setups in \emph{a)} and
\emph{b)} are equivalent, while the setup in $\emph{c)}$ has more
freedom and includes \emph{b)}. Consequently $J_a=J_b\leq J_c$.}
\end{figure}

{\noindent\emph{Proof of 2}: Notice that for any $(n+m)$-mode QFI $\mc J(\mc
S^{\star m}|\rho)$ ($m\geq n$) there exists an $(n+m+m')$-mode pure state $\ket\psi$ such that $\mc J(\mc S^{\star (m+m')}|\ket\psi)\geq \mc J(\mc S^{\star m}|\rho)$. To see this,
one simply needs to construct a pure $(n+m+m')$-mode Gaussian state
$\ket\psi$ such that $\tr_{\mc C}\ket\psi\bra\psi=\rho$, with $\mc C$
denoting the \emph{purifying} extra ($m'$) modes
[Fig.~\ref{fig:purification}\emph{b}]. Since $\mc S^{\star m} \rho=\tr_{\mc C} \mc S^{\star(m+m')}\ket\psi\bra\psi$ the monotonicity of the QFI under CPTP maps~\cite{petz_96} guarantees that
\begin{equation} \label{eq:purification}
	\mc J(\mc S^{\star (m+m')}|\ket\psi)\geq \mc J(\mc S^{\star m}|\rho).
\end{equation}
Notice that the constraint $\phi(\rho)$ is not affected by this
construction, as long as it depends only on the reduced state $\rho_{\mc
A}=\tr_{\bar{\mc A}}\rho$, since $\phi(\rho)=\phi(\ket\psi)$.
This shows that the optimal probe states can always be taken to be pure,
provided that one enlarges sufficiently the set of ancillary modes.
Therefore $C_{\pure}(\mc S^{\star (m+m')}|\phi^\star) \succeq C(\mc
S^{\star m}|\phi^\star)$ yielding $M(C_{\pure}(\mc S^{\star
(m+m')}|\phi^\star))\subseteq M(C(\mc S^{\star m}|\phi^\star))$.

Using Eq.~\eqref{eq:pureequal} we can, furthermore, say that
$M(C_{\pure}(\mc S^{\star m}|\phi^\star))\subseteq M(C(\mc
S^{\star m}|\phi^\star))$. On the other hand, it is trivial that
$C_{\pure}(\mc S^{\star m}|\phi^\star) \subseteq C(\mc S^{\star
m}|\phi^\star)$, yielding $M(C(\mc S^{\star m}|\phi^\star))\subseteq
M(C_{\pure}(\mc S^{\star m}|\phi^\star))$. Consequently
\begin{equation}
    M(C_{\pure}(\mc S^{\star m}|\phi^\star))= M(C(\mc S^{\star
m}|\phi^\star)).
\end{equation}
This concludes the proof.~$\blacksquare$\\*
}

This result is also a Gaussian version of previously known results for the estimation of finite-dimensional channels~\cite{fujiwara_identification_2001}. The added value of this result is that the restriction to pure states can be made preserving the Gaussian character of the probe states as well as respecting the resource budget $\phi^\star$.

All these results imply that $M(C(\mc S^{\star
(n+2n')}|\phi^\star))=M(C_{\pure}(\mc S^{\star
(n+2n')}|\phi^\star))=M(C_{\pure}(\mc S^{\star n}|\phi^\star))$,
so that Eq.~\eqref{eq:extendedmetric} reduces to
\begin{equation}
    \mathfrak J(\mc S\otimes \mc I)=\underset{\{j\in
M(C_{\pure}(\mc S^{\star n}|\phi^\star))\}}{\arg \inf} \det
j=\mathfrak J(\mc S).
\end{equation}

{Summarizing, we have shown that any Gaussian state of $2n$ modes (or more)
can be reduced to a $2n$-mode Gaussian \emph{pure} state which performs equally well or better than the original. The stability property thus follows.}\\*

\subsection{Computing $\mathfrak J(X)$}\label{sec:channelmetric}
We now turn to the problem of computing the metric $\mathfrak J(X)$. Take $\mc S$ as the channel for which the metric needs to be computed, under the constraint $\phi\leq\phi^\star$. Let $C^\star=C({\mc S^{\star n}}|\phi^\star)$. The problem can be written as
\begin{eqnarray}
	\textrm{minimize}& &\det j\\
	\textrm{subject to}& & m\geq \mc J\quad \forall \mc J\in C^\star.
\end{eqnarray}
This can be cast as a semi-infinite programming problem~\cite{semiinfinite}, where there are finitely many variables but infinitely many inequalities to satisfy. In the following we provide an approximation method by discretizing the problem into a convergent sequence of convex programming problems. Despite the apparently untractable nature of the problem, consistent approximations can be computed from the following prescription: Generate a finite subset of constraints by sampling a subset $C_n\subset C^\star$ of $n$ random QFI matrices from $C^\star$. Let $\mathfrak J_n$ be the solution to the discretized problem with matrices in $C_n$. Then $\mathfrak J_n$ can be computed following standard convex optimization methods. In fact, the problem of computing $\mathfrak J_n$ can be recast as the problem of finding the L\"owner-Jones ellipsoid of a union of ellipsoids~\cite{boyd_convex}~(see Fig.~\ref{fig:ellipses}), for which efficient methods exist.

We show in Appendix~\ref{app:convergence} that this approximation method converges to the true value $\mathfrak J$. More precisely, we show that for any $\epsilon>0$ there exists a sufficiently large $n$ such that $\Pr(\|\mathfrak J_n-\mathfrak J\|>\epsilon)<k\exp(-n)$, where $\|\cdot\|$ is the operator norm.

This iterative method assumes that one can generate any QFI matrix in $C^\star$. This is indeed a nontrivial
task. In principle, QFI matrices can always be computed numerically for finite dimensional quantum systems, but in the case of infinite dimensions, a general method does not exist. In the following section we concentrate on a particular class of channels and provide closed analytic formulae for the QFI matrix.

\begin{figure}
\includegraphics[width=.45\textwidth]{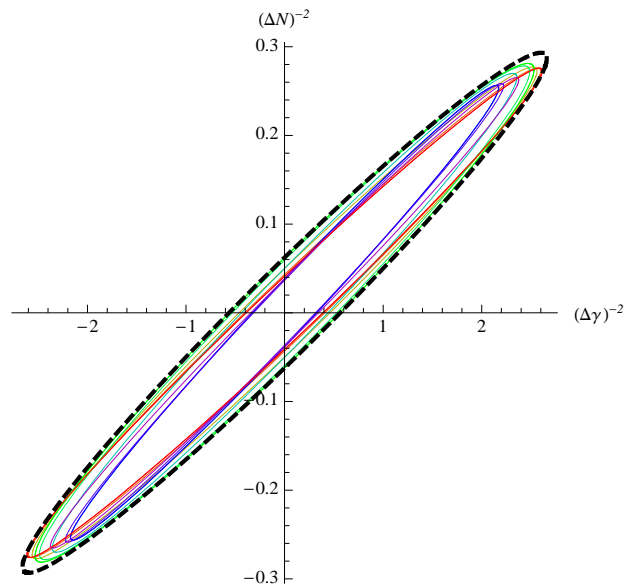}
\caption{\label{fig:ellipses}(Color online) Illustration of the problem for the model $X=(\gamma,N,0,0)$ at the point $(\gamma,N)=(0.1,1)$, tested with Gaussian states fulfilling $\phi(\rho)=\tr[a^\dagger a \,\rho]=0.2$. Ellipses correspond to the points $X \mc J^{-1} X=1$. The available energy is distributed among single mode squeezing (red), two-mode squeezing (green) and displacement (blue). Ellipsoids corresponding to the QFI's for different probe states are drawn with matching RGB color scheme. The thick-dashed ellipse corresponds to the metric $\mathfrak J$.}
\end{figure}

\section{Explicit formulae for dissipative channels}
\label{sec:formulae}

We dedicate this section to the issue of measurability and computability stated in the introduction. In principle the SLD is defined for any possible model $\mc S(X)\rho$, and thus the parameters $X$ are always measurable. Similarly, the QFI is always defined and can be obtained from the second derivatives of the quantum fidelity $F(\rho(X),\rho(X'))$ w.r.t. $X'$ at $X'=X$~\cite{hubner_explicit_1992}. Therefore, the matrix $\mc J(X)$ is always defined. However, computing the QFI in the general case can be extremely challenging. Generic formulae can be found in the literature, but obtaining explicit expressions often requires knowledge of the eigenbasis of $\mc S(X)\rho$. On the other hand, brute force evaluation of the SLD amounts to solving a Sylvester equation~\cite{bhatia_matrix_1996}. Several algorithms exist for performing such a task numerically. Unfortunately, this is of little use in the case of infinite dimensions.

In this section we develop explicit formulae for the computation of the SLD and the QFI
of dissipative channels probed by arbitrary Gaussian probe states. {Dissipative Gaussian
channels are of the form}
\begin{equation}\label{eq:channel1}
	\mc S(X)=\exp\mc G(X),
\end{equation}
where
\begin{align}\nonumber
	\mc G(X)=\sum_{k=1}^{n}\frac{\gamma_k}{2}\left(N_kL[a_k^\dagger]+(N_k+1)L[a_k]\right.\\ \label{eq:channel2}
	\left.+M_k^*D[a_k]+M_kD[a_k^\dagger]\right),
\end{align}
with $L[o]\rho=2o\rho o^\dagger-o^\dagger o \rho-\rho o^\dagger o$ and $D[o]=2o\rho o-o^2\rho-\rho o^2$ and $a^\dagger_k$, $a_k$ are creation and annihilation operators in mode $k$, satisfying the commutation relations $[a_i,a^\dagger_{j}]=\delta_{ij}$. We have used the shorthand notation $X\equiv \bigoplus_{k=1}^nx_k=(\gamma_1,N_1,\RM_1,\IM_1,\gamma_2,\ldots)$ while $x_k$ stands for $x_k=(0,\ldots,0,\gamma_k,N_k,\RM_k,\IM_k,0,\ldots,0)$, with nonzero parameter values only in mode $k$. We have chosen to work with $(\RM,\IM)$ rather than $(M,M^*)$ to ensure hermiticity of the SLD. These are the most general Gaussian dissipative channels, where the $\{\gamma,\ldots,\gamma_k\}$ parameters can be regarded as coupling strengths, $N$ corresponds to the mean photon number of the reservoir modes and the $M$'s are their squeezing parameters. Notice that this parameterization can be taken beyond the natural interpretation of the parameters, \emph{e.g.}, by setting $\gamma=\mathscr{C}-\mathscr{A}$ and $N=\mathscr{A}/(\mathscr{C}-\mathscr{A})$ one can account for amplification processes where $\mathscr{A}$, $\mathscr{B}$ and $\mathscr{C}$ are the gain, saturation and decay rate respectively in the linear regime ($\mathscr{B}=0$)~\cite{scully_quantum_optics}. Moreover, the squeezing parameters $M$ can accommodate phase sensitivity in the amplification process.

The dynamics of an arbitrary state $\rho$ undergoing the most general Gaussian dissipative evolution, in the interaction picture and within the Markovian approximation, can be described by the master equation~\cite{barnett_methods_2003}
\begin{align}
	\frac{d\rho}{dt}=\mc G(X)\rho,
\end{align}
and evolves, after a time $t$, from state $\rho_0$ to $\rho=\exp\left(t\,\mc G(X)\right)\rho_0$. For all practical purposes the coupling constants always appear as $\gamma_k t$, thus the time $t$ can be absorbed in the coupling constants $\gamma_k$, allowing to take $t=1$ without loss of generality.

There are several ways to compute the QFI. As said above, one consists in computing the Hessian of the fidelity. This, however, does not provide information about the optimal observables, namely, the SLD's $\Lambda_\mu$'s. Hence, we take here the longer route, by first computing the SLD's. This settles the issue about measurability, as we explicitly provide the optimal observables to estimate the value of $X^\mu$. Then the QFI is immediately given by their covariance matrix. In order to preserve the clarity, we introduce notation by reviewing some well-known facts about Gaussian states.

At variance with the previous sections, where we have made explicit the distinction between the channel modes $\mc A$ and the ancillary modes $\mc B, \mc C$, we now treat all modes indistinctly. Considering channels with ancillas only amounts to considering parameter spaces of the form
\begin{equation}
	X=\bigoplus_{k=1}^{n-m}\,x_k\oplus\vec0\,,
\end{equation}
where $\vec0={\bigoplus_{k=1}^m}(0,0,0,0)$. We thus we have ${n}$ bosonic modes with annihilation operators $a_k$ fulfilling the canonical commutation relations $[a_i,a_j^\dagger]=\delta_{ij}$, all other commutators being zero. We arrange all operators into a vector $\chi=(a_1,a_1^\dagger,a_2,a_2^\dagger,\ldots)$. The commutation relations are expressed as
\begin{equation}
	[\chi^i,\chi^j]=\Omega^{ij},
\end{equation}
where $\Omega=\bigoplus_{k=1}^{{n}} \omega$, and $\omega$ is the symplectic matrix
\begin{equation}
	\omega=\left(\begin{array}{cc}0 & 1 \\-1 & 0\end{array}\right),\quad\omega^T=-\omega,\quad\omega^2=-\openone.
\end{equation}
Equivalently, the canonical operators $Q_k$ and $P_k$ are arranged in the vector $R=(Q_1,P_1,Q_2,P_2,\ldots)$, related to $\chi$ by $R^i=H^i_{~j}\chi^j$ where $H$ is the unitary transformation
\begin{equation}
	H=\frac{1}{\sqrt2}\bigoplus_{k=1}^{{n}}\left(\begin{array}{cc}1 & 1 \\-i & i\end{array}\right),
\end{equation}
and with commutation relations $[R^i,R^j]=i\Omega^{ij}=H^i_{~i'}H^j_{~j'}\Omega^{i'j'}$.

A Gaussian state is defined as a state with Gaussian characteristic function~\cite{puri_mathematical_2001}, which is fully described by the first and second statistical moments,
\begin{subequations}
\begin{eqnarray}
	\langle R^i\rangle&=&\tr[\rho\,R^i],\\
	\Sigma^{ij}&=&\tr[\rho\,(R^i-\langle R^i\rangle)\circ(R^j-\langle R^j\rangle)].
\end{eqnarray}
\end{subequations}
Williamson's theorem~\cite{williamson_1936} ensures that a Gaussian state can always be expressed as a thermal state under the action of a symplectic transformation $S$ and a displacement operator $D$,
\begin{equation}
	\label{eq:williamson}
	\rho=DS \left(\bigotimes_{k=1}^n\rho_{\nu_k}\right) S^\dagger D^\dagger
\end{equation}
with
\begin{equation}
	\label{eq:thermal}
	\rho_{\nu_k}=Z_k^{-1}\exp\left(-\beta_k\, a^\dagger_ka_k\right),
\end{equation}
where $Z_k$ is a normalization factor, and the $\beta_k$ are inverse temperatures. The thermal state has zero first moments $\langle R^i\rangle=0$, and a covariance matrix $\Sigma_\therm^{ij}=\tr[(\bigotimes_{k=1}^n\rho_{\nu_k}) \, R^i\circ R^j]$,
\begin{equation}
	\Sigma_\therm=\frac12\bigoplus_{k=1}^n \nu_k\openone_2,\quad\nu_k=\coth\left(\beta_k/2\right).
\end{equation}

In order to deal with states with nonzero first moments, it is convenient to define the displaced bosonic operators $\tilde\chi^i=D\chi^i D^\dagger=\chi^i-\langle\chi^i\rangle$. The displaced canonical operators $\tilde R^i$ are defined likewise. With this, the CM reads
\begin{eqnarray}\label{eq:covariance}
	\Sigma^{ij}&=&\tr[\rho\,\tilde R^i\circ \tilde R^j].
\end{eqnarray}

Symplectic transformations are linear in the creation/annihilation operators and preserve the commutation relations,
\begin{subequations}
\begin{eqnarray}
	\label{eq:udefinition}
	S^\dagger\chi^i S&=&u^i_{~j}\chi^j,\\
\label{eq:symplectic1}
	u^i_{~i'}\,u^j_{~j'}\,\Omega^{i'j'}&=&\Omega^{ij},
\end{eqnarray}
\end{subequations}
where $u$ is a matrix representation of the symplectic transformation $S$.
The latter equation is often written as
\begin{equation}
	\label{eq:symplectic2}
	u\,\Omega \,u^T=\Omega,
\end{equation}
and we will use the fact that
\begin{equation}
	\label{eq:symplectic3}
	u^T\,\Omega=\Omega \,u^{-1}.
\end{equation}
In analogy with \eqref{eq:udefinition}, the canonical operators transform according to
\begin{equation}
	\label{eq:transfR}
	S^\dagger R^i S=H^i_{~j}u^j_{~k}[H^\dagger]^k_{~l} R^l\equiv s^i_{\,j}R^j,
\end{equation}
where we have defined $s=HuH^\dagger$, fullfilling {the same relations \eqref{eq:symplectic2} and \eqref{eq:symplectic3}. From \eqref{eq:covariance} and \eqref{eq:transfR}, the transformation} rule for the CM under a symplectic transformation $\rho\rightarrow U\rho U^\dagger$ takes the explicit form
\begin{subequations}
\begin{eqnarray}
\label{eq:transf}
	\Sigma &\stackrel{U}{\longrightarrow}&s\,\Sigma \,s^T,\\
\label{eq:transf2}
	\Sigma^{ij}&\stackrel{U}{\longrightarrow}&s^i_{\,i'}s^j_{\,j'}\Sigma^{i'j'},
\end{eqnarray}
\end{subequations}
and from Eqs.~\eqref{eq:williamson} and \eqref{eq:transf} we have $\Sigma=s\,\Sigma_\therm\,s^T$.

It is worth stressing at this point the doubly-contravariant character of the covariance matrix $\Sigma$. Although the matrix equation \eqref{eq:transf} suggests otherwise, it is clear from Eq.~\eqref{eq:transf2} that both indices transform in the same way. In order to construct functions of the covariance matrices with proper transformation rules, we will find it suitable to define the covariant-contravariant version, by lowering one index with the symplectic matrix~\cite{holevo_evaluating_2001}. Let us introduce the symplectic matrix with lower indices $\Omega_{ij}\equiv\Omega^{ij}$, so that $\Omega^{ij}\Omega_{jk}=-\delta^i_{\,k}$. Define the 1-1 \emph{covariance tensor} $\tilde\Sigma^i_{\,j}$ as
\begin{equation}
	\tilde\Sigma=\Sigma\Omega, \textrm{~~~or}\qquad \tilde\Sigma^i_{\,j}=\Sigma^{ik}\Omega_{kj}.
\end{equation}
With this definition we see that, at variance with Eqs.~\eqref{eq:transf} and \eqref{eq:transf2}, under a symplectic transformation,
\begin{eqnarray}
	\tilde\Sigma &\stackrel{U}{\longrightarrow}&s\,\tilde\Sigma \,s^{-1}.
\end{eqnarray}
Products of 1-1 tensors enjoy proper transformation rules,
\begin{eqnarray}
	[A B]^i_{~k}&=&A^i_{~j}B^j_{~k},\\
	s(AB)s^{-1}&=& (sAs^{-1})\,(sBs^{-1}).
\end{eqnarray}
Tensor products $A\otimes B$ are 2-2 tensors with indices
\begin{equation}
	[A\otimes B]^{ij}_{~~kl}=A^i_{~k}B^j_{~l}.
\end{equation}
Moreover powers of a $k$-$k$ tensor transform in the same way as $T$,
\begin{eqnarray}
	[T^n]^{i_1\ldots i_k}_{~j_1\ldots j_k}&=&T^{i_1\ldots i_k}_{~r'_1\ldots r'_k}\,T^{r'_1\ldots r'_k}_{~r''_1\ldots r''_k}\cdots T^{r^{(n-1)}_1\ldots r^{(n-1)}_k}_{~j_1\ldots j_k}\\
	(s^{\otimes k}T(s^{-1})^{\otimes k})^n&=&s^{\otimes k}T^n(s^{-1})^{\otimes k}.
\end{eqnarray}
The identity $k$-$k$ tensor is $\openone^{\otimes k}$, with $[\openone]^i_{~j}=\delta^i_{~j}$.
Finally the inverse of a tensor $T$ is defined as
\begin{equation}
T^{-1}\equiv\sum_m(\openone^{\otimes k}-T)^m
\end{equation}
Due to the antisymmetry of $\Omega$ it is important to be consistent in the way that indices are lowered and risen. Indices are lowered by contracting with the first index of $\Omega_{ij}$ and risen by contracting with the second index of $\Omega^{ij}$, so that $\Sigma^{ij}=\tilde\Sigma^i_{\,k}\Omega^{jk}=\Sigma^{il}\Omega_{lk}\Omega^{jk}=\Sigma^{il}\delta^j_l$.
%

%

%

Finally, given a probe state $\rho_0$, the action of the channel is
\begin{equation}
	\rho(X)=\mc S(X)\rho_0,\qquad	\mc S(X)=\exp\mc G(X),
\end{equation}
and the derivatives of $\rho$ w.r.t.~the channel parameters $X^\mu$ can be neatly expressed as
\begin{eqnarray}
	\label{eq:capitalD}
	\partial_\mu\rho(X)=\mc D_{\mu}\rho(X),
\end{eqnarray}
where $\mc D_\mu$ are superoperators whose expressions are given in the Appendix \ref{sec:ApI}.


\subsection{Symmetric Logarithmic Derivatives}
\label{sec:measure}
From now on we consistently drop the dependency on $X$ and assume throughout that we are considering a particular point $\mc S(X)\in\set$. The common structure of the $\mc D$ superoperators,
\begin{equation}
	\mc D_\mu\rho=\alpha_{\mu, i j} (\chi^i \rho\chi^j-(\chi^j\chi^i)\circ\rho).
\end{equation}
comes in very handy for computing the SLD's in a general manner. From Eqs.~\eqref{eq:SLDdef}  and \eqref{eq:capitalD}, these satisfy the equation,
\begin{equation}
	\label{eq:SLDeq}
	\mc D_\mu\rho=\Lambda_\mu\circ\rho.
\end{equation}
Using $\chi^j\chi^i=\chi^i\circ\chi^j-\frac{1}{2}\Omega^{ij}$ we have
\begin{equation}
	\mc D_\mu\rho=\alpha_{\mu, i j} \left[\chi^i\rho\chi^j-\left(\chi^i\circ\chi^j-\frac{1}{2}\Omega^{ij}\right)\circ\rho\right].
\end{equation}
Combining this with Eq.~\eqref{eq:SLDeq} we get
\begin{equation}
	\alpha_{\mu, i j} \chi^i\rho\chi^j=\left(\Lambda_\mu+\alpha_{\mu, i j} (\chi^i\circ\chi^j-\frac{1}{2}\Omega^{ij})\right)\circ\rho.
\end{equation}
This is a particular form of the Sylvester equation~\cite{bhatia_matrix_1996} $Y=Z \circ \rho$, which, for $\rho>0$ has the formal solution
\begin{equation}\label{eq:sylvester}
	Z=2\int_0^\infty e^{-v \rho}Y e^{-v \rho}\,dv.
\end{equation}
Therefore
\begin{align}
	\label{eq:SLDformula}
	&\Lambda_\mu=\\
	\nonumber
	&\alpha_{\mu,ij}\left[2\int_0^\infty e^{-v\rho}\, \chi^i \rho\,\chi^j\, e^{-v\rho}\,dv-\big(\chi^i\circ\chi^j-\tfrac{1}{2}\Omega^{ij}\big)\right].
\end{align}
We show in Appendix \ref{app:Integral} that
\begin{align}
\nonumber
	\int_0^\infty& e^{-u\rho}\chi^i \rho\,\chi^j\, e^{-u\rho}\,du\\
\nonumber
		=&~[H^\dagger\otimes H^\dagger]^{ij}_{~~i'j'}\times\\
		\Bigg(&\left[
			(f(\tilde\Sigma)\otimes\openone+\openone\otimes f(\tilde\Sigma)^{-1})^{-1}
		\right]^{i'j'}_{~~~kl}
\nonumber
		\Big(
			\tilde R^k\circ \tilde R^l+\frac{i}{2}\Omega^{kl}
		\Big)\\
\nonumber
		&+\big[(\openone+f(\tilde\Sigma))^{-1}\big]^{i'}_{~i''}\tilde R^{i''}\langle R^{j'}\rangle\\
\nonumber
		&+\big[(\openone+f(\tilde\Sigma)^{-1})^{-1}\big]^{j'}_{~j''}\langle R^{i'} \rangle\tilde R^{j''}\\
	\label{eq:integral}
		&+\frac{1}{2}\langle R^{i'}\rangle\langle R^{j'}\rangle\Bigg)
\end{align}
%
where $\tilde\Sigma=\Sigma\Omega$ is the 1-1 covariance tensor of $\rho$ and
\begin{equation}
	f(x)=\frac{x-i/2}{x+i/2}.
\end{equation}
Finally, defining
\begin{equation}
	\tilde\alpha_{\mu,ij}=\alpha_{\mu,i'j'}[H^\dagger\otimes H^\dagger]^{i'j'}_{~~ij},
\end{equation}
allows to express the SLD as
\begin{align}
\nonumber
	\Lambda_\mu=
	\tilde\alpha_{\mu,ij}\Big(
		\big[L^{(0)}\big]&^{ij}_{~~~kl}\Omega^{kl}+\big[L^{(1)}\big]^{ij}_{~~k}\tilde R^{k}\\
	\label{eq:SLD}
		&+\big[L^{(2)}\big]^{ij}_{~~~kl}~
		\tilde R^{k}\circ \tilde R^{l}
	\Big),
\end{align}
where

\begin{subequations}\label{eq:L}
\begin{align}
\label{eq:L0}
	L^{(0)}=&\,
	i
	\Big(
		[f(\tilde\Sigma)\otimes\openone+\openone\otimes f(-\tilde\Sigma)
	]^{-1}
	+\frac{1}{2}\openone\otimes\openone
	\Big),\\
\nonumber
	\big[L^{(1)}\big]^{ij}_{~k}=&\,\big[2(\openone+f(\tilde\Sigma))^{-1}-\openone\big]^{i}_{~k}\langle R^{j}\rangle\\
\label{eq:L1}
	&\,+\big[2(\openone+f(\tilde\Sigma)^{-1})^{-1}-\openone\big]^{j}_{~k}\langle R^{i}\rangle,\\
\label{eq:L2}
	L^{(2)}=&\,
		2
		\Big(
			[f(\tilde\Sigma)\otimes\openone+\openone\otimes f(-\tilde\Sigma)]^{-1}-\frac{1}{2}\openone\otimes\openone
		\Big).
\end{align}
\end{subequations}
Here we have used the fact that $f(x)^{-1}=f(-x)$.

Notice that this expression overcomes the main difficulty in evaluating the QFI for Gaussian channels of continuous variable systems, namely, expressing the SLD in a manageable form. In our case, the problem is reduced from working in an infinite-dimensional Hilbert space, to dealing with finite dimensional vector spaces, where the tensors $L^{(0)}$, $L^{(1)}$ and $L^{(2)}$ are defined. More importantly, these tensors depend exclusively on the covariance matrix and the first moments of the quantum state output from the channel, which have a simple relation to the channel parameters~\cite{serafini_quantifying_2005}. This will not only provide a means for evaluating the channel metric $\mathfrak J$, but it will also allow {to evaluate the performance} of several channel measurement schemes~\cite{monras_illuminati_prep}.

We conclude this subsection by commenting on possible difficulties when computing SLD's for singular states. This situation can arise when the CM of $\rho$ in Williamson form contains a vacuum mode. This is the case, for instance, when probing a zero-temperature channel ($N=0$) with a two-mode squeezed vacuum. In general the SLD is not defined on the kernel of $\rho$, \emph{i.e.}, $\overline P\Lambda\overline P$ is undetermined by Eq.~\eqref{eq:SLDdef}, where $\overline P$ is the projector on $\ker \rho$.
{Let $\mc P[\Lambda]=\Lambda-\overline P\Lambda \overline P$. By measuring $\Lambda_\nu$ on the state $\rho(X+dX)=\rho(X)+\rho(X)\circ\Lambda_\mu dX^\mu$ it is easy to see that $\tr[\rho(X+dX) \Lambda_\nu]=\tr[\rho(X+dX) \mc P[\Lambda_\nu]]$. This is the freedom available in defining SLD's for singular $\rho$'s. However, Eqs.~\eqref{eq:SLD} and \eqref{eq:L} were derived from Eq.~\eqref{eq:SLDeq} by means of expression \eqref{eq:sylvester} which assumes that the density operator is nonsingular. Thus, these expressions can yield to divergencies for $\overline P\Lambda_\mu\overline P$. These are not observable divergencies, and can be regularized by introducing a small temperature $\epsilon$ in the probe state $\rho_0$, projecting $\Lambda$ with $\mc P$, and finally taking the limit $\epsilon$ to zero.} There may be, on the other hand, observable divergencies. This situation often arises when one explores the vicinity of the boundary of the manifold. The classical Fisher information is well known to diverge in many statistical models, the most prominent case being the binomial distribution when $p\rightarrow0$ or $1$, giving rise to Poissonian statistics. {Analogously}, the Bures distance for mixed qubit states is well known to diverge in the limit of pure states, giving rise to interesting and counterintuitive effects in Bayesian qubit estimation~\cite{bagan_collective_2004}. A similar effect in quantum statistics has been reported in a particular Gaussian channel [$X=(\gamma,0,0,0)$], when the parameter $\gamma\rightarrow 0$~\cite{monras_optimal_2007}.

\subsection{The Quantum Fisher Information matrix}
The expression of the SLD, Eq.~\eqref{eq:SLD}, allows for the computation of the QFI, Eq.~\eqref{eq:QFIdef}. The full derivation of the most general QFI is given in Appendix~\ref{sec:App4}, Eq.~\eqref{eq:QFIfull}. We reproduce here the resulting expression:
\begin{align}
\label{eq:QFIsimplified}	\mc J_{\mu\nu}&(X)=
	 \tilde\alpha_{\mu,i'j'}\tilde\alpha_{\nu,k'l'}~\bigg(\big[L^{(1)}\big]^{i'j'}_{~~i}\big[L^{(1)}\big]^{k'l'}_{~~k}\Sigma^{ik}\\
	\nonumber
	 &+\big[L^{(2)}\big]^{i'j'}_{\,\,i''j''}\big[L^{(2)}\big]^{k'l'}_{\,\,kl}[D]^{i''j''}_{\,\,ij}\left(\Omega^{ik}\Omega^{jl}+\Omega^{il}\Omega^{jk}\right)\bigg).
\end{align}
It should be noted that the tensors appearing in Eq.~\eqref{eq:QFIsimplified} depend on $\tilde\Sigma$, the covariance tensor of the \emph{output} state. In order to finally evaluate the QFI one needs to express $\tilde\Sigma$ in terms of the initial quantum state $\Sigma_0$ and the channel parameters $X$. The corresponding relation is well known in the literature~\cite{serafini_multiplicativity_2005,serafini_quantifying_2005} but we repeat it here for completeness. Arranging the channel parameters $X$ in the \emph{asymptotic} covariance matrix $\Sigma_\textrm{ch}$,
\begin{equation}\label{eq:finalstateCM1}
	\Sigma_\textrm{ch}=\bigoplus_k\left({\footnotesize \begin{array}{cc}\frac{1}{2}+N_k+\Re M_k & \Im M_k \\\Im M_k  & \frac{1}{2}+N_k+\Re M_k\end{array}}\right)
\end{equation}
and defining the coupling matrix
\begin{eqnarray}\label{eq:finalstateCM2}
	\Gamma&=&\bigoplus_ke^{-\frac{\gamma_k}{2}}\openone_2,
\end{eqnarray}
we have
\begin{equation}\label{eq:finalstateCM3}
	\Sigma=\Gamma(\Sigma_0-\Sigma_{\textrm{ch}})\Gamma+\Sigma_{\textrm{ch}},
\end{equation}
where $\Sigma_0$ is the CM of the initial state, and $\Sigma$ is the CM of the state going out of the channel.

Eqs.~\eqref{eq:QFIsimplified}, \eqref{eq:finalstateCM1}, \eqref{eq:finalstateCM2} and \eqref{eq:finalstateCM3} provide a systematic way to compute the QFI for any channel $\mc S(X)$ and any input state $\rho_0$. Remarkably, this expression is \emph{exact} and \emph{analytical}. However, despite the significant simplification of the problem -- namely, from an infinite dimensional operator equation \eqref{eq:SLDeq} to a finite-dimensional matrix expression, Eq.~\eqref{eq:QFIsimplified}-- explicit analytical expressions are too complex to be of any use, except for very simple channel models.
\section{Conclusion and outlook}
\label{sec:conclusions}

Summarizing, for any resource constraint $\phi(\rho_{\mc A})\leq\phi^\star$ we have defined a metric tensor on the manifold of Gaussian channels. We have proven that the metric is stable under the addition of ancillas and provided a method for computing numerical approximations, by using convex optimization methods. The resulting distance is an upper bound to the attainable Bures distance between states resulting from the action of the channel onto any given initial Gaussian state fulfilling the constraints, and it allows to establish a systematic way to measure distances between channels in order to quantify imperfections in quantum information implementations. Moreover, the metric tensor minimizes the volume element assigned at any point of the set of Gaussian channels. This density can be used as an analog of the Jeffrey's (Bures) prior distribution for points in the simplex (density matrices) in Bayesian estimation methods. Our results are a step towards the identification of a useful notion of distance among bosonic channels, which are the basis mathematical framework of quantum communication devices with continuous variables.

Additionally, our approach has provided several results which are of relevance to the field of quantum estimation theory. We have derived closed formulae for computing the QFI of any estimation method that uses Gaussian states and dissipative Gaussian channels. Moreover, for each one of the channel parameters we have obtained closed expressions for the corresponding optimal observables associated to each one of the parameters (symmetric logarithmic derivatives). Moreover, we have proved that by enlarging sufficiently the number of ancillary modes, the optimal probe states for estimating Gaussian channels are always pure.

These findings should be of immediate use in several practical situations. The most prominent of them is entanglement-enhanced metrology of Gaussian channels. A first relevant application of our results is the evaluation of Gaussian protocols for parameter estimation in dissipative Gaussian channels, such as decay rate, temperature or degree of squeezing in engineered baths~\cite{monras_illuminati_prep}. The stability proof in Sec.~\ref{sec:stability} establishes the maximum number of ancillary modes to be considered for any of these situations. Indeed, such proof is very general and can be extended to several metrology problems, limiting the number of ancillary modes required for optimality, thus restricting the number of setups that need to be explored when designing optimal metrology protocols. In the spirit of the Choi-Jamiolkowski theorem, our arguments show that estimation of single mode channels cannot be enhanced by multipartite entanglement as compared to bipartite entanglement, at least within the Gaussian framework.

Moreover, the approach taken here may be extended to derive SLD's for other related problems such as phase estimation under decoherence, phase diffusion, etc, which can be expressed quite naturally in phase space, and that would otherwise be difficult to address due to the infinite dimensional character of continuous variable systems.

Our work leaves open several questions and possible extensions. Obviously, obtaining explicit analytic expressions for the channel metric would be the ultimate achievement. However, this seems to be out of reach unless some significant advances are made. In particular we point out three relevant missing points; \emph{a}) A clear criterion for the determination of a set of optimal probe states for any given choice of $\phi$. \emph{b}) A general form of the QFI for all Gaussian channels, not only the dissipative ones. \emph{c}) An analytic expression for the metric in the high resource limit (properly regularized). Concentrating our attention on the regularization schemes, we can point out some interesting questions: What classes of constraints $\phi$ will always be saturated by the optimizing states? Considering a related problem, preliminary numerical results~\cite{monras_illuminati_prep} show that in the single-mode bosonic lossy channel, at zero-temperature, with constraints of the form $\phi(\rho)=\tr[a^\dagger a \rho]$, an entangled two-mode squeezed state of the probe and ancilla modes with mean photon number $\langle n\rangle$ in the channel mode will perform equally well any other state with $n$ average photons. If the result of this preliminary analysis will be confirmed, it is unlikely to be a coincidence. Finding the reasons behind this surprising match is an interesting problem worth deeper investigation.

Naturally, these are not the only open questions, and in particular, extending our work to non-Gaussian channels and/or non-Gaussian probes would be of utmost importance. There is, in fact, nothing peculiar in our defining scheme that relies crucially on Gaussianity of the channels and/or the probe states. The metric could thus be defined on more general sets of channels and probe states. It is however, the restriction to Gaussian channels and states that allows us to provide computational formulae and prove that our metric satisfies the basic requirements listed in the introduction, especially the stability property. This does not mean that non-Gaussian extensions will not fulfill the requirements. Indeed, it would be an interesting result if a counterexample to stability could be found in the large set of non-Gaussian states and channels. It would mean that some estimation protocols can be improved by considering multipartite setups, when only one part (mode) undergoes the channel action. This would be an effect exclusively of continuous variable systems which cannot occur in finite dimensions.

Finally, a question remains unanswered regarding the way in which channels are combined. Throughout the text we have assumed that the channel is tested a large number $N$ of times, \emph{independently}, namely, a given probe state is prepared, sent through the channel and measured, and the process is repeated $N$ times. The number of trials is immediately regularized and thus does not enter in the discussion. However, in the quantum scenario, regularizing $N$ leaves no room for chaining channels together, so that, for instance, a state may be sent twice through the channel before being measured, or entangled states may be used in the simultaneous testing of a single and a doubly chained channel. It is evident that a large number of degrees of freedom are not exploited in our approach. However, our framework yields the most natural bounds, which can be obtained in the \emph{iid} ("independent and identically distributed") case, when no correlation is admitted between different samples. A very relevant and interesting open question is whether such sophisticated schemes can enhance the precision of quantum estimation and quantum metrology setups.\\*

\noindent \textbf{Acknowledgements.}
The authors wish to thank Dr. F. Dell'Anno, Dr. S. M. Giampaolo, and Dr. J. Virto for
discussions. They are especially indebted to Prof. A. Winter for his encouraging and
instructive support, and in particular for his inspiring suggestions regarding the
proof of convergence in Appendix~\ref{app:convergence}.
The authors acknowledge financial support from the European Commission of the
European Union under the FP7 STREP Project HIP (Hybrid Information Processing),
Grant Agreement n. 221889, from MIUR under the FARB fund, from INFN under
Iniziativa Specifica PG 62, and from CNR-INFM Center Coherentia.
One of us (F. I.) acknowledges support from the ISI Foundation
for Scientific Interchange.

\appendix
\section{Convergence of the approximation algorithm}
\label{app:convergence}

We show here that the approximation method given in Sec.~\ref{sec:channelmetric} does converge to the true value of the metric $\mathfrak J$. Throughout this appendix we make extensive use of the Hausdorff distance $d_H(\cdot,\cdot)$~\cite{munkres_topology_2000}, which defines a distance among subsets of a metric space $W$. In our case $W$ is the set of positive semidefinite matrices, equipped with the operator norm distance $d(x,y)=\|x-y\|$, where $\|x\|$ is the operator norm $\|x\|=\sup_{|\vec v|=1}|x \vec v|$. Define the $\epsilon$-ball centered at a point $x$ as the set of points within a distance $\epsilon$ of $x$, $B_\epsilon(x)=\{x'~|~d(x,x')<\epsilon\}$. Then, the $\epsilon$-neighborhood of a subset $X$ is the union of all $\epsilon$-balls of $X$, $B_\epsilon(X)=\cup_{x\in X}B_\epsilon(x)$. Then, the Hausdorff distance is defined as
\begin{equation}
	d_H(X,Y)=\inf\{\delta~|~X\subset B_\delta(Y)~\textrm{and}~Y\subset B_\delta(X)\}.
\end{equation}

Let us first show some preliminary facts about the randomly generated sets of constraints, $C_n$.
\\*

\noindent\textbf{Theorem:}  Assuming that the set $C^\star=C(\mc S|\phi^\star)$ of QFI matrices associated to a given channel $\mc S$ [within the corresponding probe state restrictions $\phi(\rho)\leq \phi^\star$] is bounded, we have the following property: For any $\epsilon >0$ the probability $\Pr(d_H(C_n,C^\star)\geq\epsilon)$ decreases exponentially.\\*

\noindent \proof: Since the set $C^\star$ is bounded it can be covered by a finite number $k$ of $\epsilon/2$-balls centered at points $\{y_i\in C^\star\},~i=1\ldots,k$, thus $C^\star\subseteq B_{\epsilon/2}(\{y_i\})$. If
\begin{equation}\label{eq:cover}
	B_{\epsilon/2}(\{y_i\})\subset B_{\epsilon}(C_n)
\end{equation}
then $C^\star\subseteq B_{\epsilon/2}(\{y_i\})\subset B_{\epsilon}(C_n)$. Since also $C_n\subset C^\star\subset B_\epsilon(C^\star)$ we have that Eq.~\eqref{eq:cover} implies $d_H(C_n,C^\star)\leq\epsilon$.

We now compute an upper bound to the probability that Eq.~\eqref{eq:cover} is true.
To each random selection $x\in C_n$ we assign the point $y\in\{y_i\}$ which is closest to $x$. Since $d(x,y)<\epsilon/2$ then $B_{\epsilon/2}(y)\subseteq B_\epsilon(x)$. Let $n_i$ be the number of points of $C_n$ assigned to ball $B_{\epsilon/2}(y_i)$. A sufficient condition for Eq.~\eqref{eq:cover} to hold is $n_i\neq0~\forall i$ [all points in $\{y\}$ have been assigned \emph{at least} one point $x\in C_n$]. Thus
\begin{equation}
	n_i\neq0~\forall i~~\Longrightarrow~~ d_H(C_n,C^\star)\leq\epsilon.
\end{equation}
Consequently,
\begin{align}
	\Pr(d_H(C_n,C^\star)\leq\epsilon)~\geq~ \Pr(n_i\neq0~\forall i).
\end{align}
Let  $p_i$ be the probability that a random sample $x$ is assigned to point $y_i$. The probability that after $n$ samplings a number $n_1,\ldots,n_k$ of points has been assigned to each ball is given by the multinomial distribution
$$
	\Pr(n_1,\ldots,n_k)=\frac{n!}{n_1!\cdots n_k!}p_1^{n_1}\cdots p_k^{n_k}
$$

We now bound the complementary probabilities
\begin{align}\nonumber
\Pr(d_H(C_n,C^\star)&\,>\epsilon)\leq \Pr(\exists i ~s.t.~n_i=0)\\
\nonumber
	=&\sum_{n_2,\ldots,n_k\neq0}\Pr(0,n_2,\ldots,n_k) + perm.\\
\nonumber
	&+\sum_{n_3,\ldots,n_k\neq0}\Pr(0,0,\ldots,n_k) + perm.\\
\nonumber
	&+\ldots\\
	&+\Pr(0,0,\ldots,0,n) + perm.	,
\end{align}
where \emph{perm.} represents all permutations among arguments of the multinomial distribution and it is implicit that summation is over all $n_k$ values such that $\sum_k n_k=n$. We can now upper bound $\Pr(\exists i ~s.t.~n_i=0)$ by completing the sums to include indices equal to zero, and by using the multinomial theorem.
\begin{align}\nonumber
	\Pr(\exists i ~s.t.~&n_i=0)\\
\nonumber
	\leq&~(p_2+p_3+\ldots p_k)^n+ perm.\\
\nonumber
	&+(p_3+\ldots+p_k)^n + perm.\\
\nonumber
	&+\ldots\\
\nonumber
	&+p_k^n + perm.\\	
\nonumber
	=&~(1-p_1)^n+ perm.\\
\nonumber
	&+(1-p_1-p_2)^n + perm.\\
	&+\ldots\\
\nonumber
	&+(1-p_1-p_2-\ldots-p_{k-1})^n + perm.\,,
\end{align}
To conclude we can further upper bound this quantity by replacing each term $(\cdot)^n$ by the maximum value $(1-\min_i p_i)^n$ and counting the number of terms $\sum_{l=1}^{k-1}\binom{k}{l}=2^k-2$,
\begin{align}
	\Pr(\exists i ~s.t.~n_i=0)\leq (2^k-2) (1-\min_i p_i)^n
\end{align}

Notice that this is exponentially decreasing with $n$. Thus, we have that for any $\epsilon>0$ the probability $\Pr(d_H(C_n,C^\star)\,>\epsilon)$ decreases exponentially with $n$. This means that by picking enough many samples we have $d_H(C_n,C^\star)\leq\epsilon$  with arbitrarily high probability.$~\blacksquare$\\*

Let $x\in X$ be a positive semidefinite matrix within a set $X$. Define $M(x)$ as the set of all upper bounds to $x$, and $M(X)$ as the set of upper bounds to all elements in $X$, \emph{i.e.}, $M(X)=\{m\,|\,m\geq x~\forall x\in X\}$. We call $M(X)$ the set of \emph{feasible points} to the optimization problem.\\*

\noindent{\bf Lemma:} Given two matrices $x$ and $y$ such that $d(x,y)<\delta$ it holds that $x-\delta\openone<y<x+\delta\openone$.\\*

\noindent\proof: Define $\Delta=y-x$. Then $\|\Delta\|=d(x,y)<\delta$. Moreover $-\|\Delta\|\openone\leq \Delta\leq\|\Delta\|\openone$ so that $-\delta\openone<\Delta<\delta\openone$. Thus,
\begin{eqnarray}
	x-\delta\openone<y<x+\delta\openone.
\end{eqnarray}
\begin{flushright}
	{$\blacksquare$}
\end{flushright}

\noindent{\bf Theorem:} The map $M:X\mapsto M(X)$ is continuous in the Hausdorff distance. Namely, for any $\epsilon>0$ there exists a $\delta>0$ such that $d_H(X,Y)<\delta~\Rightarrow~d_H(M(X),M(X))< \epsilon$.\\*

\noindent\proof: For any $\epsilon>0$ take $\delta<\epsilon$. We prove constructively by showing that if $d_H(X,Y)<\delta$ it follows that $M(X)\subset B_\epsilon(M(Y))$ and $M(Y)\subset B_\epsilon(M(X))$ which in turn implies that $d_H(M(X),M(Y))<\epsilon$.

We have to show that $a\in B_\epsilon(M(Y))~\forall a\in M(X)$. Take any element $a\in M(X)$ and construct $b=a+\delta\openone$. It must follow that (\emph{i}) $b\in M(Y)$ and (\emph{ii}) $d(a,b)<\epsilon$.
(\emph{i}) Since $d_H(X,Y)<\delta$ then for all $y\in Y$ we have an $x\in X$ such that $d(x,y)< \delta$ and by the lemma, $x+\delta\openone > y$. Thus, $b=a+\delta\openone>x+\delta\openone>y$ for all $y\in Y$. Thus $b\in M(Y)$. (\emph{ii}) $d(a,b)=\delta<\epsilon$. This proves that $M(X)\subset B_\epsilon(M(Y))$. The exact same reasoning with $X$ and $Y$ interchanged shows the converse, thus $d_H(M(X),M(Y))<\epsilon$.$~\blacksquare$\\*

The last two theorems combined show that given any $\epsilon>0$, one can always generate a set of feasible points $M(C_n)$ which has Hausdorff distance from $M(C^\star)$ smaller than $\epsilon$ with arbitrarily high probability, \emph{i.e.}, $\Pr(d_H(M(C_n),M(C^\star))<\epsilon)\geq1-k\exp(-n)$. Define the function
\begin{equation}
	f(X)=\underset{x\in X}{\arg\inf}\det x.
\end{equation}
It only remains to prove that $\mathfrak J_n=f(M(C_n))$ approaches $\mathfrak J$ as $M(C_n)$ approaches $M(C^\star)$. Let us remark that, by construction $M(C^\star)\subset M(C_n)$. As said in the text, we assume that the solution to the real problem $\mathfrak J=f(M(C^\star))$ is nondegenerate. This means that $\det(\mathfrak J+\Delta)>\det\mathfrak J$ whenever $\mathfrak J+\Delta\in M(C^\star)$ and $\|\Delta\|\neq0$.\\*

\noindent{\bf Theorem}: Assuming that the miminum of $\{\det y~|~y\in Y\}$ is nondegenerate, the map $\{f(X_n)\}$, converges to $f(Y)$ in the Hausdorff distance whenever $\{X_n|X_n\supset Y~\forall n\}$ converges to $Y$. Namely, for any $\epsilon>0$ there is a $\delta>0$ such that if $Y\subset X$ and $d_H(X,Y)<\delta$ then $\|f(X)-f(Y)\|<\epsilon$.\\*

\noindent\proof: Let $x^\star=f(X)$ and $y^\star=f(Y)$. 
Consider an open ball of radius $\dmax<\epsilon/2$ around $y^\star$, and let $\Gamma$ be the smallest gap between determinants of points in $Y$ outside $B_{\dmax}(y^\star)$ and  $\det y^\star$, \emph{i.e.},
\begin{equation}
	\Gamma=\inf_{\substack{y\notin B_{\dmax}(y^\star)\\y\in Y}}\det y-\det y^\star.
\end{equation}
By the nondegeneracy assumption $\Gamma>0$. Consider the function
\begin{equation}
	h(\delta)=n(\|y^\star\|+D+\delta)^{n-1}\delta
\end{equation}
where $D$ is the maximum distance between points in the boundary of $Y$. Clearly $h$ is monotonically increasing and $h(0)=0$. Let
\begin{equation}
	\delta<\min\left(h^{-1}(\Gamma),\epsilon/2\right),
\end{equation}

By assumption $X\subset B_\delta(Y)$ and $Y\subset X$. Let $y\in Y$ be the closest point to $x^\star$. It is easy to see that $y$ must lie in the boundary of $Y$, since if it was not, there would be a ball $B_\tau(y)\subset Y$ and $y'=y-\tau\|x^\star-y\|\openone\in B_\tau(y)\subset Y$ would have $\|y'-x^\star\|=(1-\tau)\|y-x^\star\|$, which contradicts the assumption that $y$ is the closest point to $x^\star$ in $Y$.

 We know that $\|x^\star -y\|<\delta$. Let $y=y^\star +\Delta$. By the triangle inequality
\begin{equation}\label{eq:triangle}
	\|x^\star-y^\star\|\leq \|\Delta\|+\delta.
\end{equation}
Also, since $Y\subset X$ we have $\det x^\star\leq \det y^\star$, thus
\begin{equation}\label{eq:detineq}
	\det y-\det y^\star\leq \det y-\det x^\star.
\end{equation}
Moreover (see \cite{bhatia_matrix_1996})
\begin{eqnarray}\label{eq:detbound}
	\det y-\det x^*\leq n\max(\|y\|,\|x^\star\|)^{n-1}\|x^\star -y\|.
\end{eqnarray}
We can further simplify this by noticing that $\|x^\star\|=\|y+(x^\star-y)\|\leq\|y\|+\|x^\star-y\|\leq\|y\|+\delta$, thus $\max(\|y\|,\|x^\star\|)\leq\|y\|+\delta\leq\|y^\star\|+\|\Delta\|+\delta$, so we have
\begin{equation}
	\det y-\det x^*\leq n(\|y^\star\|+\|\Delta\|+\delta)^{n-1}\delta.
\end{equation}
Notice that $\|\Delta\|=\|y-y^\star\|$ is the distance between two points in the boundary of $Y$, thus $\|\Delta\|\leq D$, so that
\begin{eqnarray}\label{eq:detbound2}
	\det y-\det x^*\leq h(\delta).
\end{eqnarray}

On the other hand, let us show by contradiction that $\|\Delta\|<\dmax$. Suppose $y\notin B_{\dmax}(y^\star)$. By Eq.~\eqref{eq:detbound2}, the monotonicity of $h$ and the definition of $\delta$ we have
\begin{eqnarray}
	\det y-\det x^\star\leq h(\delta)<\Gamma.
\end{eqnarray}
which means that $\det x^\star>\det y-\Gamma$. But $y\notin B_{\dmax}(y^\star)$ means $\det y-\det y^\star\geq \Gamma$, which yields $\det x^\star>\det y^\star$. This contradiction shows that $\|\Delta\|\leq\dmax$, which in turn shows that $\|\Delta\|<\epsilon/2$.

Finally, recovering Eq.~\eqref{eq:triangle} we obtain
\begin{equation}
	\|x^\star-y^\star\|<\epsilon.
\end{equation}
\begin{flushright}$\blacksquare$
\end{flushright}

Taking $\mathfrak J_n=f(M(C_n))$ and $\mathfrak J=f(M(C^\star))$ and applying this theorem to $f(M(C_n))$ and $f(M(C^\star))$, where $C_n\subset C^\star$ and $M(C^\star)\subseteq M(C)$ with
\begin{equation}
	\Pr(d_H(M(C_n),M(C^\star))>\delta)\stackrel{n\rightarrow\infty}\longrightarrow0
\end{equation}
 will ensure that
\begin{equation}
	\Pr(\| \mathfrak J_n-\mathfrak J\|>\epsilon)\stackrel{n\rightarrow\infty}\longrightarrow0
\end{equation}

  Since the probability to generate a $C_n$ within any given Hausdorff distance to $C^\star$ approaches 1 exponentially, we see that any desired precision can be attained with arbitrarily high certainty by sampling enough matrices from $C^\star$.

\section{Derivaties w.r.t. the channel parameters}
\label{sec:ApI}
In order to compute the derivatives of $\rho(X)=\mc S(X)\rho_0$, note that
\begin{equation}
	\partial_\mu\rho(X)=\partial_\mu\mc S(X)\,\rho_0.
\end{equation}
Thus we only need to compute the derivative of the superoperator $\mc S(X)$.
Using the relation~\cite{wilcox_exponential_1967}
\begin{align}
\label{eq:wilcox}
	\frac{\partial}{\partial X^\mu}\exp\mc G(X)=
	\int_0^1 e^{u \mc G(X)}\partial_\mu \mc G(X) e^{(1-u) \mc G(X)}du,
\end{align}
the channel derivatives can be written as
\begin{equation}
	\partial_\mu\mc S(X)=\int_0^1 e^{u \mc G(x_\kappa)}\partial_\mu \mc G(x_\kappa) e^{-u \mc G(x_\kappa)}du \, \mc S(X),
\end{equation}
where $\kappa$ stands for the mode to which the parameter $X^\mu$ corresponds. This expression allows for the handy relation
\begin{equation}
	\partial_\mu\mc S(X)=\mc D_\mu \, \mc S(X),
\end{equation}
with
\begin{equation}
	\mc D_\mu=\int_0^1 e^{u \mc G(x_\kappa)}\partial_\mu \mc G(x_\kappa) e^{-u \mc G(x_\kappa)}du.
\end{equation}

Using the Hadamard lemma,
\begin{equation}
	\exp(uA) B \exp(-uA)=\sum_{m=0}^\infty\frac{u^m}{m!}[A,B]_m,
\end{equation}
with $[A,B]_0\equiv B$ and $[A,B]_m\equiv [A,[A,B]_{m-1}]$, and the commutation relations
\begin{subequations}
\begin{eqnarray}
	\big[L[a_k^\dagger],L[a_{k'}]\big]&=&2\delta_{kk'}\big(L[a_k]+L[a_k^\dagger]\big),\\
	\big[D[a_k^\dagger],D[a_{k'}]\big]&=&0,\\
	\big[D[a_k],L[a_{k'}]\big]&=&2\delta_{kk'}D[a_k],\\
	\big[L[a_k^\dagger],D[a_{k'}]\big]&=&2\delta_{kk'}D[a_k],\\
	\big[D[a_{k}^\dagger],L[a_{k'}]\big]&=&2\delta_{kk'}D[a_k^\dagger],\\
	\big[L[a_k^\dagger],D[a_{k'}^\dagger]\big]&=&2\delta_{kk'}D[a_k^\dagger],
\end{eqnarray}
\end{subequations}
one can show by induction that
\begin{subequations}
\begin{eqnarray}
	\big[\mc G(X),L[a_k]\big]_m&=&\gamma_k^m L[a_k]+2 \gamma_k^{m-1}\mc G(x_k),\\
	\big[\mc G(X),L[a_k^\dagger]\big]_m&=&\gamma_k^m L[a_k^\dagger]-2 \gamma_k^{m-1}\mc G(x_k),\\
	\big[\mc G(X),D[a_k]\big]_m&=&\gamma_k^m D[a_k],\\
	\big[\mc G(X),D[a_k^\dagger]\big]_m&=&\gamma_k^m D[a_k^\dagger],
\end{eqnarray}
\end{subequations}
%
and the $\mc D$ superoperators become
\begin{subequations}
\begin{eqnarray}
	\mc D_{\gamma_k}&=&\frac{1}{\gamma_k} \mc G(x_k),\\
	\mc D_{N_k}&=&(e^{\gamma_k}-1)\frac{L[a_k]+L[a_k^\dagger]}{2},\\
	\mc D_{\RM_k}&=&(e^{\gamma_k}-1)\frac{D[a_k]+D[a_k^\dagger]}{2},\\
	\mc D_{\IM_k}&=&(e^{\gamma_k}-1)\frac{D[a_k]-D[a_k^\dagger]}{2i}.
\end{eqnarray}
\end{subequations}
Finally, notice that the $\mc D$ superoperators allow for evaluating derivatives of the state $\rho$ in a simple form,
\begin{equation}
	\partial_\mu\rho=\partial_\mu S(X) \rho_0=\mc D_\mu S(X) \rho_0=\mc D_\mu \rho,
\end{equation}
where all have the structure
\begin{equation}
	\mc D_\mu \rho=\alpha_{\mu,ij}(\chi^i \rho\chi^j-(\chi^j\chi^i)\circ\rho).
\end{equation}
\section{Computing the SLD}
\label{app:Integral}
We proceed here to show how Eqs.~\eqref{eq:integral}, \eqref{eq:SLD} and \eqref{eq:L} are derived. We remark that the relations obtained apply to all Gaussian states in any number of modes. We use the notations introduced in Sec.~\ref{sec:formulae}.\\*

\noindent\textbf{Lemma:} $[\rho_\therm,\chi^i]=[\exp M-\openone]^i_{\,j}\chi^j \rho_\therm$ where $M=\bigoplus_k\textrm{diag}(\beta_k,-\beta_k)$.

\noindent\proof: We begin by showing
\begin{eqnarray}
	\rho_\therm \chi^i \rho_\therm^{-1}=[\exp{M}]^i_{\,j}\chi^j,
\end{eqnarray}
Define $A=-\bigoplus_k\beta_k a_k^\dagger a_k$ and $B_m=[A,B_{m-1}]$; $B_0=\chi^i$.
One can show by induction that $B_m=[M^m]^i_{\,j}\chi^j$.
Using the Hadamard lemma,
\begin{eqnarray}
	\rho_\therm \chi^i \rho_\therm^{-1}&=&\sum_m \frac{1}{m!}B_m=[\exp{M}]^i_{\,j}\chi^j,
\end{eqnarray}
thus, finally
\begin{align}\nonumber
	[\rho_\therm,\chi^i]&=\left(\rho_\therm\chi^i \rho_\therm^{-1}-\chi^i\right)\rho_\therm\\
	&=[\exp M-\openone]^i_{\,j}\chi^j \rho_\therm~\blacksquare
\end{align}
\noindent\textbf{Lemma:} Let $u$ be the symplectic representation of $S$ as defined in Eq.~\eqref{eq:udefinition} and $s$ as defined in Eq.~\eqref{eq:transfR}, where $S$ is defined from $\rho$ in Eq.~\eqref{eq:williamson}. Let $\Sigma$ be the covariance matrix of $\rho$ (Eq.~\eqref{eq:covariance}) and $E=\exp M$. Then
\[
	u\,E\,u^{-1}=H^\dagger f(\tilde\Sigma) H
\]
where
\[
	f(x)=\frac{x-i/2}{x+i/2}.
\]
\proof: Observe that
\begin{eqnarray}
	f(\tilde\Sigma)&=&\frac{\tilde\Sigma^2-\openone/4-i\tilde\Sigma}{\tilde\Sigma^2+\openone/4}\\
\nonumber
	 &=&s\frac{\tilde\Sigma_\therm^2-\openone/4-i\tilde\Sigma_\therm}{\tilde\Sigma_\therm^2+\openone/4}s^{-1}\\
\nonumber
	&=&s~\bigoplus_k\left(
		\begin{array}{cc}
			\cosh\beta_k & i\sinh\beta_k  \\
			-i\sinh\beta_k & \cosh\beta_k \\
		\end{array}
	\right)s^{-1}\\
\nonumber
	&=&s H E H^\dagger s^{-1}\\
\nonumber
	&=&HuEu^{-1}H^\dagger~\blacksquare
\end{eqnarray}

\noindent\textbf{Lemma:} Given $\rho>0$ with CM $\Sigma$, define $F=H^\dagger f(\tilde\Sigma)H$.
The following relation holds
\[
	e^{-v\rho}\tilde\chi^ie^{v\rho}=\sum_m\frac{(-v)^m}{m!}[(F-\openone)^m]^i_{\,i'}\,\tilde\chi^{i'}\rho^m.
\]
\noindent\proof:
Again, using the Hadamard lemma,
\begin{equation}
	e^{-v\rho_\therm}\chi^ie^{v\rho_\therm}=\sum_m\frac{(-v)^m}{m!}C_m
\end{equation}
where $C_0=\chi^i$ and $C_m=[\rho_\therm,C_{m-1}]$. One can show by induction that $C_m=[(E-\openone)^m]^i_{\,j}\chi^j \rho_\therm^m$
. Now observe that
\begin{eqnarray}
\nonumber
	e^{-v\rho}\tilde\chi^ie^{v\rho}&=&DSe^{-v\rho_\therm}S^\dagger\chi^iS e^{v\rho_\therm}S^\dagger D^\dagger\\
\nonumber
&=&s^i_{~i'}~DSe^{-v\rho_\therm}\chi^{i'} e^{v\rho_\therm}S^\dagger D^\dagger\\
\nonumber
	&=&\sum_m\frac{(-v)^m}{m!}s^i_{~i'}[(E-\openone)^{m}]^{i'}_{~i''}~DS\chi^{i''}\rho_\therm^m S^\dagger D^\dagger\\
\nonumber
	 &=&\sum_m\frac{(-v)^m}{m!}u^i_{~i'}[(E-\openone)^{m}]^{i'}_{~i''}[u^{-1}]^{i''}_{~i'''}~\tilde\chi^{i'''}\rho^m\\
	&=&\sum_m\frac{(-v)^m}{m!}[(F-\openone)^m]^i_{~i'}\,
	\tilde\chi^{i'}\rho^m.~\blacksquare
\end{eqnarray}

\noindent\textbf{Lemma:} Given $\rho>0$ with CM $\Sigma$ and $\tilde \chi^i=D\chi^i D^\dagger$, the following holds
\begin{align}\nonumber
	\int_0^\infty e^{-v\rho}\tilde \chi^ie^{-v\rho}dv=&\big[(\openone+F)^{-1}\big]^i_{~i'}\tilde\chi^{i'}\rho^{-1}\\
	=&\big[(\openone+F^{-1})^{-1}\big]^i_{~i'}\rho^{-1}\tilde\chi^{i'}
\end{align}
\emph{Proof}:
\begin{align}\nonumber
	\int_0^\infty&e^{-v\rho}\tilde \chi^ie^{-v\rho}dv\\
\nonumber
	=&\sum_m\big[(F-\openone)^m\big]^{i}_{i'}\tilde\chi^{i'}\rho^m \int_0^\infty dv\frac{(-v)^m}{m!}e^{-2v\rho}\\
\nonumber
	=&\sum_m\frac{1}{2}\left[\left(\frac{\openone-F}{2}\right)^m\right]^i_{~i'}\tilde \chi^{i'}\rho^{-1}\\
	=&\big[(\openone+F)^{-1}\big]^i_{~i'}\tilde\chi^{i'}\rho^{-1}.
\end{align}
The other identity is proven analogously~$\blacksquare$\\*

\noindent\textbf{Theorem:} Given $\rho>0$, the following relation holds
\[
	\int_0^\infty e^{-v\rho}\, \tilde\chi^i \rho\,\tilde\chi^j\, e^{-v\rho}\,dv=[(F\otimes\openone+\openone\otimes F^{-1})^{-1}]^{ij}_{\,\,\,\,kl}\,\tilde\chi^k\tilde\chi^l,
\]
where the inverse of a tensor $T^{ij}_{\,\,kl}$ is $T^{-1}$ such that
\[
	T^{ij}_{\,\,kl}[T^{-1}]^{kl}_{\,\,rs}=\delta^i_{\,r}\delta^j_{\,s}.
\]
\proof: Making extensive use of the previous lemma,
\begin{align}
\nonumber
	\int_0^\infty& e^{-v\rho}\, \tilde\chi^i \rho\,\tilde\chi^j\, e^{-v\rho}\,dv\\
\nonumber
	=&\sum_m\frac{1}{m!}[(F-\openone)^m]^i_{~k}\,\tilde\chi^k\\
\nonumber
	 &\times~\int_0^\infty (-v)^m\rho^{m+1} e^{-v\rho}\,\tilde\chi^j\, e^{-v\rho}\,dv\\
\nonumber
	 =&\sum_{mn}\frac{(-1)^m}{m!n!}[(F-\openone)^m]^i_{\,k}[(F-\openone)^n]^j_{\,l}\,\tilde\chi^k\rho^{1+m+n}\\
\nonumber
	&\times~\int_0^\infty v^{m+n} e^{-2v\rho}\,\rho^{-n}\tilde\chi^l\rho^n\,du\\
\nonumber
	=&\sum_{mn}\frac{(-1)^m(m+n)!}{2^{1+m+n}m!n!}[(F-\openone)^m]^i_{\,k}[(F-\openone)^n]^j_{\,l}\,\\
\nonumber
	&\times~\tilde\chi^k\rho^{-n}\tilde\chi^l\rho^n\\
\nonumber
	=&\sum_{mn}\frac{(-1)^m}{2^{1+m+n}}\left(\begin{array}{c}m+n \\n\end{array}\right)[(F-\openone)^m]^i_{\,k}[(F-\openone)^n]^j_{\,l}\\
\nonumber
	&\times~[F^{-n}]^l_{\,s}\,\tilde\chi^k\tilde\chi^s\\
\nonumber
	=&\sum_{mn}\frac{(-1)^m}{2^{1+m+n}}\left(\begin{array}{c}m+n \\n\end{array}\right)[(F-\openone)^m]^i_{\,k}[(\openone-F^{-1})^n]^j_{\,l}\\
	&\times~\tilde\chi^k\tilde\chi^l.
\end{align}
Next, make the change of variables $m+n=q$ and replace $\sum_{m,n=0}^\infty$ by $
\sum_{q=0}^\infty\sum_{n=0}^q$,
\begin{align}\nonumber
	&\int_0^\infty e^{-v\rho}\, \tilde\chi^i \rho\,\tilde\chi^j\, e^{-v\rho}\,dv\\
\nonumber
	&=\sum_{q=0}^\infty\frac{(-1)^q}{2^{q+1}}\sum_{n=0}^q
	\left(\begin{array}{c}q \\n	\end{array}\right)
	[(F-\openone)^{q-n}\otimes(F^{-1}-\openone)^n]^{ij}_{~~kl}\\
	&=\left[\left(F\otimes\openone+\openone\otimes F^{-1}\right)^{-1}\right]^{ij}_{~~kl}.
\end{align}
This proves the theorem. $\blacksquare$\\*

Finally we are in position to obtain Eq.~\eqref{eq:integral}. Observe that
\begin{align}
	\int_0^\infty&e^{-v\rho}\, \chi^i \rho\,\chi^j\, e^{-v\rho}\,dv\\
\nonumber
	=&\int_0^\infty e^{-v\rho}\, (\tilde\chi+\langle \chi\rangle)^i \rho\,(\tilde\chi+\langle \chi\rangle)^j\, e^{-v\rho}\,dv.
\end{align}
Collecting the results and using
\begin{align}
	[H\otimes H]^{kl}_{~~k'l'}~\tilde\chi^{k'}\tilde\chi^{l'}&= \tilde R^k \tilde R^l\\ \nonumber
			&=\tilde R^k\circ \tilde R^l+\frac{i}{2}\Omega^{kl},\\
	[H\otimes H]^{kl}_{~~k'l'}~\tilde\chi^{k'}\langle\chi^{l'}\rangle&=\tilde R^{k}\langle R^{l}\rangle\\
	[H\otimes H]^{kl}_{~~k'l'}~\langle\chi^{k'}\rangle\langle\chi^{l'}\rangle&=\langle R^{k}\rangle\langle R^{l}\rangle
\end{align}
we finally obtain
\begin{align}
\nonumber
	\int_0^\infty& e^{-u\rho}\chi^i \rho\,\chi^j\, e^{-u\rho}\,du\\
\nonumber
		=&~[H^\dagger\otimes H^\dagger]^{ij}_{~~i'j'}\times\\
		\Bigg(&\left[
			(f(\tilde\Sigma)\otimes\openone+\openone\otimes f(\tilde\Sigma)^{-1})^{-1}
		\right]^{i'j'}_{~~~kl}
\nonumber
		\Big(
			\tilde R^k\circ \tilde R^l+\frac{i}{2}\Omega^{kl}
		\Big)\\
\nonumber
		&+\big[(\openone+f(\tilde\Sigma))^{-1}\big]^{i'}_{~i''}\tilde R^{i''}\langle R^{j'}\rangle\\
\nonumber
		&+\big[(\openone+f(\tilde\Sigma)^{-1})^{-1}\big]^{j'}_{~j''}\langle R^{i'} \rangle\tilde R^{j''}\\
		&+\frac{1}{2}\langle R^{i'}\rangle\langle R^{j'}\rangle\Bigg)
\end{align}
which is Eq.~\eqref{eq:integral}.

\section{Some simplifying identities}
\label{sec:App3}
In this section we derive some relations that will be useful for the simplification and evaluation of the QFI. Consider the tensor $[f(\tilde\Sigma)\otimes\openone+\openone\otimes f(-\tilde\Sigma)]^{-1}$. Notice that if $[X,Y]=0$ we can write
\begin{align}
	 \big[f(X)+f(Y)\big]^{-1}=\frac{1}{2}\left(\frac{XY-\openone/4}{XY+\openone/4}+\frac{i}{2}\frac{(X+Y)}{XY+\openone/4}\right).
\end{align}
Given that $f(X)\otimes\openone=f(X\otimes\openone)$, $\openone\otimes f(Y)=f(\openone\otimes Y)$ and $[\tilde\Sigma\otimes\openone,\openone\otimes\tilde\Sigma]=0$ we can write
\begin{align}
	\big[f(\tilde\Sigma)&\otimes\openone+\openone\otimes f(-\tilde\Sigma)\big]^{-1}
	\\&=\frac{1}{2}
\nonumber	D^{-1}\left(
		 \tilde\Sigma\otimes\tilde\Sigma+\frac{1}{4}\openone\otimes\openone-\frac{i}{2}(\tilde\Sigma\otimes\openone-\openone\otimes\tilde\Sigma)\right)
\end{align}
where we have defined
\begin{equation}
	\label{eq:Ddef}
	D=\tilde\Sigma\otimes\tilde\Sigma-\frac{1}{4}\openone\otimes\openone,
\end{equation}
thus
\begin{align}
	 L^{(0)}&=i\,D^{-1}\Big(\tilde\Sigma\otimes\tilde\Sigma-\frac{i}{4}(\tilde\Sigma\otimes\openone-\openone\otimes\tilde\Sigma)\Big),\\
	L^{(2)}&=\frac{1}{2}D^{-1}\big(
			\openone\otimes\openone-i(\tilde\Sigma\otimes\openone-\openone\otimes\tilde\Sigma)
		\big).
\end{align}

At this point, the following relations are useful,
\begin{subequations}
\begin{align}
\nonumber
	 \big[\tilde\Sigma\otimes\tilde\Sigma\big]^{ij}_{~~kl}\Omega^{kl}&=[\Sigma\Omega\Omega(\Sigma\Omega)^T]^{ij}\\
	&=[\Sigma\Omega\Sigma]^{ij},\\
\nonumber
	 \big[\tilde\Sigma\otimes\openone-\openone\otimes\tilde\Sigma\big]^{ij}_{~~kl}\Omega^{kl}&=[\Sigma\Omega\Omega-\Omega(\Sigma\Omega)^T]^{ij},\\
	&=-2\Sigma^{ij}\\
\nonumber
	 \big[\tilde\Sigma\otimes\tilde\Sigma\big]^{ij}_{~~kl}\Sigma^{kl}&=[\Sigma\Omega\Sigma(\Sigma\Omega)^T]^{ij}\\
	&=-[\Sigma\Omega\Sigma\Omega\Sigma]^{ij},\\
\nonumber
	 \big[\tilde\Sigma\otimes\openone-\openone\otimes\tilde\Sigma\big]^{ij}_{~~kl}\Sigma^{kl}&=[\Sigma\Omega\Sigma-\Sigma(\Sigma\Omega)^T]^{ij}\\
	&=2[\Sigma\Omega\Sigma]^{ij}.
\end{align}
\end{subequations}
Thus finally we get
\begin{eqnarray}
	 \big[L^{(0)}\big]^{ij}_{~~kl}\Omega^{kl}&=&i\left[D^{-1}\left(\Sigma\Omega\Sigma+\frac{i}{2}\Sigma\right)\right]^{ij},\\
	 \big[L^{(2)}\big]^{ij}_{~~kl}\Sigma^{kl}&=&-i\left[D^{-1}\left(\Sigma\Omega\Sigma+\frac{i}{2}\Sigma\right)\right]^{ij},
\end{eqnarray}
thus
\begin{equation}
	\big[L^{(0)}\big]^{ij}_{~~kl}\Omega^{kl}=-\big[L^{(2)}\big]^{ij}_{~~kl}\Sigma^{kl}.
	\label{eq:consistency2}
\end{equation}
This identity, together with the fact that $\tr[\rho \tilde R^i]=0$ guarantees the consistency check $\tr[\rho \Lambda_\mu]=0$, Eq.~\eqref{eq:consistency1}. It also turns out to be very handy in the simplification of the QFI, in Appendix~\ref{sec:App4}.

\section{Computing the QFI}\label{sec:App4}
The QFI can be computed from Eqs.~\eqref{eq:QFIdef}, \eqref{eq:SLD} and \eqref{eq:L}. Combining Eqs.~\eqref{eq:QFIdef} and \eqref{eq:SLD} we obtain all possible cross terms among $L^{(0)}$, $L^{(1)}$ and $L^{(2)}$. Classifying terms according to the number of $\tilde R$ operators they contain, we find the following. Of zero order only one term is obtained~\eqref{J:zero}. Terms in first order contain products of $L^{(0)}$ and $L^{(1)}$. These terms vanish when the expectation is taken ($\langle \tilde R\rangle=0$). Second order terms contain cross products between $L^{(0)}$ and $L^{(2)}$ and "second powers" of $L^{(1)}$, lines~\eqref{J:second1}, \eqref{J:second2} and \eqref{J:second3}. Third order terms contain expressions of the form $\tr[\rho \tilde R^i(\tilde R^j\circ\tilde R^k)]$, which vanish identically [we argue this claim at the end of the appendix]. There is only one fourth order term \eqref{J:fourth}, which comes from double contribution of $L^{(2)}$.
\begin{subequations}\label{J:golbal}
\begin{align}\nonumber
	\mc J_{\mu\nu}&(X)=
	\tilde\alpha_{\mu,ij}\tilde\alpha_{\nu,kl}~\times\\
\label{J:zero}
	\Big(\,&\big[L^{(0)}\big]^{ij}_{\,\,i'j'}\Omega^{i'j'}\big[L^{(0)}\big]^{kl}_{\,\,k'l'}\Omega^{k'l'}\\
\label{J:second1}
	+&\big[L^{(1)}\big]^{ij}_{\,\,i'}\big[L^{(1)}\big]^{kl}_{\,\,k'}\tr[\rho \tilde R^{i'}\circ\tilde R^{k'}]\\
\label{J:second2}
	+&\big[L^{(0)}\big]^{ij}_{\,\,i'j'}\Omega^{i'j'}\big[L^{(2)}\big]^{kl}_{\,\,k'l'}\Sigma^{k'l'}\\
\label{J:second3}
	+&\big[L^{(2)}\big]^{ij}_{\,\,i'j'}\Sigma^{i'j'}\big[L^{(0)}\big]^{kl}_{\,\,k'l'}\Omega^{k'l'}\\
\label{J:fourth}
	+&\big[L^{(2)}\big]^{ij}_{\,\,i'j'}\big[L^{(2)}\big]^{kl}_{\,\,k'l'}\tr[\rho (R^{i'}\circ R^{j'})		 \circ(R^{k'}\circ R^{l'})]
	~\Big).
\end{align}
\end{subequations}
Using Eq.~\eqref{eq:consistency2} this is reduced to
\begin{align}\label{eq:QFI2}
	\mc J_{\mu\nu}&(X)=~
	\tilde\alpha_{\mu,ij}\tilde\alpha_{\nu,kl}~\times \nonumber \\
	\bigg(&\big[L^{(1)}\big]^{ij}_{\,\,i'}\big[L^{(1)}\big]^{kl}_{\,\,k'}\Sigma^{i'k'}
	+\big[L^{(2)}\big]^{ij}_{\,\,i'j'}\Sigma^{i'j'}\big[L^{(0)}\big]^{kl}_{\,\,k'l'}\Omega^{k'l'}
\nonumber \\
	&+\big[L^{(2)}\big]^{ij}_{\,\,i'j'}\big[L^{(2)}\big]^{kl}_{\,\,k'l'}\tr[\rho (R^{i'}\circ R^{j'})		 \circ(R^{k'}\circ R^{l'})]
	\bigg).
\end{align}
At this point a manageable expression for $T^{ijkl}=\tr[\rho\, (R^{i}\circ R^{j})\circ(R^{k}\circ R^{l})]$ is an imperative. This tensor is symmetric under interchange of indices (12), (34) and simultaneous interchange (13)(24), which generate a subgroup $\mathfrak S$ of the full symmetric group $S_4$. It is convenient to establish the relation between the tensor $T^{ijkl}$ and the fully symmetric one
\begin{equation}
	T_W^{ijkl}=\frac{1}{4!}\sum_{\sigma\in S_4}T^{\sigma(ijkl)},
\end{equation}
which can be easily computed with standard phase space methods~\cite{barnett_methods_2003},
\begin{eqnarray}
	T_W^{ijkl}=\tr[\rho\, W(R^iR^jR^kR^l)],
\end{eqnarray}
where $W(X)$ is the Weyl-ordered product of $X$. Being $\Sigma$ the covariance matrix of $\rho$, $T_W^{ijkl}$ is just the fourth moment of the Gaussian distribution with covariance matrix $\Sigma$, namely,
\begin{eqnarray}
	T_W^{ijkl}=\Sigma^{ij}\Sigma^{kl}+\Sigma^{ik}\Sigma^{jl}+\Sigma^{il}\Sigma^{jk}.
\end{eqnarray}

In order to exploit the preexisting symmetry of $T$ we split the sum over $\sigma\in S_4$ over right cosets $S_4/\mathfrak S$, of which we pick $\{e,(23),(24)\}$ as representatives, namely $\mathfrak S\cup (23)\mathfrak S\cup(24)\mathfrak S= S_4$. We can rewrite $T_W$ as
\begin{align}
	T_W^{ijkl}=\frac{1}{4!}\sum_{a\in\{e,(23),(24)\}}\sum_{\sigma\in \mathfrak S}T^{a\sigma(ijkl)}.
\end{align}
Since $T$ is invariant under the action of $\mathfrak S$ we have
\begin{align}
	T_W^{ijkl}&=\frac{|\mathfrak S|}{4!}\sum_{a\in\{e,(23),(24)\}}T^{a(ijkl)} \nonumber \\
	&=\frac{1}{3}\left(T^{ijkl}+T^{ikjl}+T^{ilkj}\right).
\end{align}
Therefore
\begin{eqnarray}
\nonumber
	 T^{ijkl}&=&T_W^{ijkl}+\frac{1}{3}\left(T^{ijkl}-T^{ikjl}\right)+\frac{1}{3}\left(T^{ijkl}-T^{ilkj}
\right) \nonumber \\
	&=&T_W^{ijkl}+\frac{2}{3}\left(T_{[2,3]}^{ijkl}+T_{[2,4]}^{ijkl}\right) \, ,
\end{eqnarray}
where $T_{[2,3]}^{ijkl}$ is the antisymmetrized tensor in the second and third indices,
\begin{align}
	T_{[2,3]}^{ijkl}&=\frac{1}{2}\left(T^{ijkl}-T^{ikjl}\right) \nonumber \\
	&=\frac{1}{8}\left(\Omega^{ij}\Omega^{kl}-\Omega^{ik}\Omega^{jl}\right)
-\frac{1}{4}\Omega^{il}\Omega^{jk} \, ,
\end{align}
and $T_{[2,4]}^{ijkl}=\frac{1}{3}\left(T^{ijkl}-T^{ilkj}\right)=\frac{1}{3}\left(T^{ijlk}-T^{iljk}\right)=T_{[2,3]}^{ijlk}$. With this we obtain
\begin{align}
	T^{ijkl}=T_W^{ijkl}-\frac{1}{4}\left(\Omega^{ik}\Omega^{jl}+\Omega^{il}\Omega^{jk}\right).
\end{align}
Thus finally
\begin{align}
	T^{ijkl}=\Sigma^{ij}\Sigma^{kl}\,+\,&\Sigma^{ik}\Sigma^{jl}+\Sigma^{il}\Sigma^{jk}%
	\nonumber \\
	-\frac{1}{4}\big(&\Omega^{ik}\Omega^{jl}+\Omega^{il}\Omega^{jk}\big).
\end{align}
It will be convenient to write this result as
\begin{eqnarray}\label{eq:T}
	 T^{ijkl}=\Sigma^{ij}\Sigma^{kl}+[D]^{ij}_{\,\,i'j'}\left(\Omega^{i'k}\Omega^{j'l}
+\Omega^{i'l}\Omega^{j'k}\right) ,
\end{eqnarray}
where $D$ is defined in Eq.~\eqref{eq:Ddef}.

A similar approach shows that $\tilde R^i\circ(\tilde R^j\circ \tilde R^k)=W(\tilde R^i \tilde R^j \tilde R^k)\,+$ terms containing only first powers of $\tilde R$. It is easy to see that expectation values of all these terms vanish identically.

Using the expression for $T^{ijkl}$ we get
\begin{align}
\label{eq:QFIfull}
	\mc J_{\mu\nu}&(X)=
	 \tilde\alpha_{\mu,i'j'}\tilde\alpha_{\nu,k'l'}~\bigg(\big[L^{(1)}\big]^{i'j'}_{~~i}\big[L^{(1)}
\big]^{k'l'}_{~~k}\Sigma^{ik} \nonumber \\
	 &+\big[L^{(2)}\big]^{i'j'}_{\,\,i''j''}\big[L^{(2)}\big]^{k'l'}_{\,\,kl}[D]^{i''j''}_{\,\,ij}\left(
\Omega^{ik}\Omega^{jl}+\Omega^{il}\Omega^{jk}\right)\bigg) \, ,
\end{align}
where, by virtue of Eq.~\eqref{eq:consistency2}, the second term in Eq.~\eqref{eq:QFI2} has been canceled against the first contrubition coming from Eq.~\eqref{eq:T}.

\end{document}